\documentclass[iop]{emulateapj}

\shorttitle{Chondrule Accretion in Oligarchic growth}
\shortauthors{Matsumoto et al.}

\begin{document}

\title{Chondrule Accretion with a Growing Protoplanet}

\author{Yuji Matsumoto,\altaffilmark{1,2}
		Shoichi Oshino,\altaffilmark{2}
		Yasuhiro Hasegawa, \altaffilmark{3}
		\& Shigeru Wakita\altaffilmark{2}
		}

\email{ymatsumoto@cfca.nao.ac.jp}

\altaffiltext{1}{Planetary Exploration Research Center, Chiba Institute of Technology, Narashino, Chiba, 275-0016, Japan}
\altaffiltext{2}{Center for Computational Astrophysics, National Astronomical Observatory of Japan, Osawa, Mitaka, Tokyo, 181-8588, Japan}
\altaffiltext{3}{Jet Propulsion Laboratory, California Institute of Technology, Pasadena, CA 91109, USA}

\begin{abstract}

Chondrules are primitive materials in the Solar System. 
They are formed in the first about 3 Myr of the Solar System's history.
This timescale is longer than that of Mars formation, 
and it is conceivable that protoplanets, planetesimals and chondrules might have existed simultaneously in the solar nebula.
Due to protoplanets¡Ç perturbation on the planetesimal dynamics and chondrule accretion on them,
 all the formed chondrules are unlikely to be accreted by planetesimals. 
We investigate the amount of chondrules accreted by planetesimals in such a condition. 
We assume that a protoplanet is in oligarchic growth, and we perform analytical calculations of chondrule accretion both by a protoplanet and by planetesimals.
Through the oligarchic growth stage, planetesimals accrete about half of the formed chondrules. 
The smallest planetesimals get the largest amount of the chondrules, compared with the amount accreted by more massive planetesimals. 
We perform a parameter study and find that this fraction is not largely changed for a wide range of parameter sets.

\end{abstract}


\section{Introduction}\label{sect:intro}

Chondrules are mm-sized spherical particles found in chondritic meteorites. 
The properties of them suggest that their precursors were melted by some flash heating events in gas nebula \citep[e.g.,][]{Scott&Krot2005, Scott2007}. 
They make up $\sim 20$ to 80 \% of most chondrites' volume,
 and their formation started from the time of Ca-Al rich inclusions (CAIs) formation and continued for at least $\sim 3$ Myr \citep{Connelly+2012}.
This means that the heating events are common in the first 3 Myr of the Solar System formation.

Several formation mechanisms of chondrules are proposed \cite[e.g.,][]{Desch+2012}.
These include X-wind model \cite[e.g.,][]{Shu+1996, Shu+2001}, nebular lightning model \citep[e.g.,][]{Desch&Cuzzi2000, Muranushi2010}, 
 nebula shock model \citep[e.g.,][]{Iida+2001, Mann+2016}, and impact jetting model \citep{Johnson+2015, Hasegawa+2016a, Hasegawa+2016b}. 
These models can reproduce some petrologic and geochemical aspects of chondrules \citep{Rubin2000}.
The models also need to explain chondrule abundance. 

The amount of chondrules, which can be inferred from chondrites, is not equal to the produced amount of chondrules. 
This is not only because the present asteroid belt mass is much smaller than that of the primordial one,
 but also because it is unclear how parent bodies of chondrites formed. 
When the currently available chondrites were generated as fragments of massive bodies \citep[e.g.,][]{Demeo+2015}, there is the following possibility \citep{Hasegawa+2016b};
 even if planetesimals originally did not contain any chondritic materials, they could accrete chondrules as chondrules formed with time. 
In this case, the planetesimals could have a chondrule-rich surface layer. 
Following the subsequent collisional cascade, such a surface layer could break into chondrites. 
In order to examine this possibility, it is important to investigate how chondrules were accreted by massive bodies such as planetesimals and protoplanets.

Recent studies \citep{Ormel&Klahr2010, LJ2012} investigated the accretion process of small particles onto massive bodies in laminar disk gas,
 known as pebble accretion.
The particles which are strongly affected by gas drag, such as chondrules, boulders, or fragments of larger bodies,
 are efficiently accreted by massive planetesimals and protoplanets.
Accretion of chondrules through pebble accretion was studied by \cite{Johansen+2015}.
They considered the situation that planetesimals are born and grow in an ocean of chondrules.
However, chondrules were formed during 3 My after CAI formation,
 and it is conceivable that planet formation actively took place at that time.
In fact, \cite{Dauphas&Pourmand2011} suggested that the timescale of Mars formation is $1.8^{+0.9}_{-1.0}$ Myr or less after CAI formation.
If such a body is in the planetesimal swarm, which can be the parent bodies of chondrites, runaway and oligarchic growth of the body occur \citep{Wetherill&Stewart1989, Kokubo&Ida1996, Kokubo&Ida1998}. 
It is therefore crucial to explore how chondrule formation and accretion occur simultaneously with the growth of protoplanets.
The accretion efficiency of chondrules by planetesimals decreases when protoplanets affect the dynamics of planetesimals \citep{Levison+2015}.
This is because planetesimals tend to be kicked out from the pebble sea due to the gravitational interaction with protoplanets,
 which increases both eccentricity and inclination of the planetesimals.
\cite{Hasegawa+2016b} studied the pebble accretion of chondrules by planetesimals, assuming that chondrules are formed by impact jetting. 
In this formation scenario, chondrule-forming impacts are realized when protoplanets are present in planetesimal disks \citep{Johnson+2015}. 
They found that there are certain ranges of parameters that satisfy the timescale of chondrule formation,
 magnetic field strength estimated from the Semarkona ordinary chondrite \citep{Fu_R+2014}, and the condition of efficient pebble accretion. 
However, the accretion efficiency of chondrules onto planetesimals and a protoplanet is not directly calculated in the previous studies.

In this paper, we investigate chondrule accretion under the presence of a growing protoplanet. 
Since the timescale of runaway growth is much smaller than that of chondrule formation,
 we consider that a protoplanet is already in the oligarchic stage, which is put in a swarm of planetesimals.
We adopt the impact jetting model as a chondrule forming process in the fiducial model.
We calculate the growth of a protoplanet analytically.
The chondrule accretion rates by a protoplanet and planetesimals are also calculated in each timestep as the protoplanet grows.
Moreover, we obtain the mass of accreted chondrules.
Our model is described in detail in section \ref{sect:model}.
In section \ref{sect:chondrule_acc}, we present the results in which the timescale of chondrule accretion
 by a protoplanet and planetesimals and the amount of chondrules accreted by them are both shown.
In section \ref{sect:discussion}, we discuss implications of chondrule accretion and physical processes that are not included in this paper.
Finally, section \ref{sect:conclusion} contains our conclusions.

\section{MODEL}\label{sect:model}

\begin{deluxetable*}{lll}
	\tablenum{1}
	\tablecaption{Summary of Key Quantities}\label{table:quantities}
	\tablehead{ 
	Symbol			&	Meanings																	&	Value 
	}
	\startdata
	$\rho_{\rm g}$		&	Gas volume density at the disk midplane											&					\\
	$f_{\rm d}$		&	Increment factor of $\rho_{\rm g}$ and $\Sigma_d$									&					\\
	$h_{\rm g}$		&	Gas pressure scale height													&					\\
	$\tau_{\rm g}$		&	Timescale of disk gas depletion												&					\\
	\hline 
	$r$				&	Orbital radius																&					\\
	$T_K$			&	Orbital period																&					\\
	$M$				&	Mass of the protoplanet														&					\\
	$\tau_{\rm pr}$		&	Timescale of the protoplanet growth 											&					\\
	$t_{\rm iso}$		&	Time until the protoplanet reaches the isolation mass								&					\\
	$M_{\rm iso}$		&	Isolation mass of the protoplanet												&					\\
	$M_{\rm esc}$		&	Mass of the protoplanet when impact velocities exceed 2.5 $\mbox{km s}^{-1}$			&					\\
	$M_{\rm ini}$		&	Mass of the protoplanet when the oligarchic growth begins							&					\\
	\hline
	$m_{\rm pl}$		&	Mass of planetesimals														&					\\
	$R_{\rm pl}$		&	Radius of planetesimals														&					\\
	$\rho_{\rm pl}$		&	Material density of planetesimals												&	2 g cm$^{-3}$		\\
	$e_{\rm pl} $		&	Eccentricity of planetesimals in oligarchic growth									&					\\
	$n_{\rm pl}$		&	Number of planetesimals														&					\\
	\hline
	$M_{\rm ch}$		&	Mass of field chondrules														&					\\
	$r_{\rm ch}$		&	Characteristic size of chondrules												&	1 mm			\\
	$\rho_{\rm s}$		&	Bulk density of chondrules													&	3.3 g cm$^{-3}$	\\
	$\rho_{\rm ch}$		&	Spatial density of chondrules in protoplanetary disk									&					\\
	$h_{\rm ch}$		&	Scale height of chondrules													&					\\
	$\tau_{\rm stop}$	&	Timescale of gas drag on chondrules											&					\\
	$F_{\rm ch}$		&	Mass fraction of planetesimals that can eventually generate chondrules via impact jetting	&	$10^{-2}	$		\\
	\hline
	$r_{\rm H}$		&	Hill radius 			 													&					\\
	$r_{\rm B}$		&	Bondi radius 																&					\\
	$M_t$			&	Transition mass 															&					\\
	$f_{\rm acc}$		&	Increment factor for chondrule accretion by planetesimals							&					\\
	$M_{\rm acc}$		&	Mass of accreted chondrules		 											&					\\
	$\tau_{\rm acc}$	&	Timescale of chondrule accretion	 											&					\\
	$\tau_{\rm B}$		&	Timescale that chondrules across $r_{\rm B}$ 										&					\\
	\hline
	$f_{{\rm r},i}$		&	The mass fraction of chondrule accreted by planetesimals in $i$-th mass bin				&					\\
	$f_{\rm m,ch}$		&	The mass fraction of chondrules with respect to an accreting planetesimal				&					\\
	$\Delta R_{\rm ch}$	&	The thickness of the chondrule layer on a planetesimal								&					
	\enddata 
\end{deluxetable*}

We introduce our models that are constituted from the combination of a disk model, chondrule formation model, and chondrule accretion model.
We consider the mass of the smallest planetesimals ($m_{\rm pl,min}$), an orbital radius ($r$), timescale of gas depletion ($\tau_g$),
 and the accretion enhancement factor ($f_{\rm acc}$) as parameters. 
In our fiducial model, 
 $m_{\rm pl,min}=10^{23}$ g planetesimals are located at $r=$2 au, the gas density is constant with time ($\tau_{\rm g}=\infty$), and $f_{\rm acc}=1$.
This set of parameters are adopted because the timescale of chondrule formation by the impact jetting process is consistent with data from chondrites \citep{Hasegawa+2016a}
 and chondrules can be accreted efficiently by planetesimals \citep{Hasegawa+2016b},
While the size of $10^{23}$ g planetesimals may apparently be (about 230 km radius with the material density of $2\mbox{ g cm}^{-3}$) too large for asteroids, 
 \cite{Morbidelli+2009} showed that the size distribution of asteroids can be reproduced when the initial planetesimals are larger than 100 km sized ones.
Table \ref{table:quantities} summarizes the important physical quantities.

\subsection{Disk model}

At first, we introduce a disk model that consists of dust and gas. 
We adopt a power-law disk model similar to the minimum-mass solar nebula model \citep{Hayashi1981}. 
Following \cite{Kokubo&Ida2000} and \cite{Hasegawa+2016a}, we give the surface density of dust ($\Sigma_{\rm d}$)
 and the surface density of gas ($\Sigma_{\rm g}$), as
\begin{eqnarray}
	\Sigma_{\rm d} &=& 10 \times f_d \left( \frac{r}{1 \mbox{ au} } \right)^{-3/2} \mbox{ g cm}^{-2}, \\
	\Sigma_{\rm g} &=& 2400 \times f_d \left( \frac{r}{1 \mbox{ au} } \right)^{-3/2} \mbox{ g cm}^{-2}, \label{eq:Sigma_g}
\end{eqnarray}
where $f_d$ is an increment factor.
In this paper, $f_{\rm d}$ is a parameter.
Reflecting the results of \cite{Hasegawa+2016a}, we consider a massive disk case, $f_d=3$, in our fiducial model.
The stellar mass is 1 solar mass. 
Under the optically thin limit, the disk temperature is given by 
\begin{eqnarray}
	T &=& 280 \left( \frac{r}{1 \mbox{ au} } \right)^{-1/2} \mbox{ K}, 
\end{eqnarray}
and the sound speed ($c_{\rm s}$), gas pressure scale height ($h_{\rm g}$), and density of gas ($\rho_{\rm g}$) are
\begin{eqnarray}
	c_{\rm s} &=& 1.1\times10^5 \left( \frac{r}{1 \mbox{ au} } \right)^{-1/4} \mbox{ cm s}^{-1}, \\
	h_{\rm g} &=& 4.7 \times 10^{-2} \left( \frac{r}{1 \mbox{ au} } \right)^{5/4} \mbox{ au},\\
	\rho_{\rm g} &=& 2\times10^{-9} f_d \left( \frac{r}{1 \mbox{ au} } \right)^{-11/4} \mbox{ g cm}^{-3}, \label{eq:rho_g}
\end{eqnarray}
respectively.
In some calculations, we consider gas depletion. 
For these calculations, the timescale of gas depletion ($\tau_g$), and $\Sigma_{\rm g}$ and $\rho_{\rm g}$
 are multiplied by $\exp{(-t/\tau_{\rm g})}$, where $t$ is time (cf. equations (\ref{eq:Sigma_g}) and (\ref{eq:rho_g})).
In disks, gas component moves with a sub-Keplerian velocity. 
The velocity can be written as $(1-\eta)v_K$, where $v_K$ is the Keplerian velocity, and 
\begin{eqnarray}
	\eta \simeq 1.8\times 10^{-3} \left( \frac{r}{1 \mbox{ au} } \right)^{1/2},
\end{eqnarray}
\citep{Nakagawa+1986}.

For the velocities of chondrules, it is determined by the degree of coupling with gas.
In this paper, we adopt 1 mm as a chondrule size, which is a typical value for chondrules found in chondrites \citep{Scott&Krot2005, Scott2007}.
Provided that chondrules are subjected to the Epstein drag force, their stopping time $\tau_{\rm stop}$ is given by 
\begin{eqnarray}
	\tau_{\rm stop} &=& \frac{\rho_{\rm s} r_{\rm ch}}{c_{\rm s} \rho_{\rm g}} \nonumber \\
		&\simeq& 5.0 \times 10^{-5} f_d^{-1} \left( \frac{r_{\rm ch}}{\mbox{1 mm}} \right) \left( \frac{\rho_{\rm s}}{\mbox{3.3 g cm}^{-3}} \right) \nonumber \\ && \times
		\left( \frac{r }{\mbox{1 au}} \right)^{3/2} T_K , \label{eq:tau_stop}
\end{eqnarray}
where $r_{\rm ch}$ is the radius of chondrules, $\rho_{\rm s}$ is the material density of them \citep{Adachi+1976, Weidenschilling1977},
 and $T_K$ is the orbital period, $T_K=2\pi/\Omega_K$, where $\Omega_K$ is a Kepler frequency.
Since the stopping time is much shorter than the orbital period, chondrules are well coupled with disk gas, and chondrules are on circular orbits.
This indicates that when chondrules were formed by impact jetting, chondrules could go out of the feeding zone of a protoplanet along with the gas motion there.

The vertical scale height of chondrules ($h_{\rm ch}$) is important for the accretion of chondrules \citep{Levison+2015}. 
Since vertical diffusion of chondrules is affected by turbulence, $h_{\rm ch}$ is determined by the strength of turbulence and $\tau_{\rm stop}$.
We use the $\alpha_{\rm eff}$ parameter to describe the strength of turbulence \citep{Shakura&Sunyaev1973}.
As suggested for protoplanetary disks, magnetic fields and the resultant disk turbulence probably played an important role for the evolution of the solar nebula. 
For this case, $\alpha_{\rm eff}$ can be written as a function of magnetic fields \citep[e.g.,][]{Wardle2007}; 
\begin{eqnarray}
	\alpha_{\rm eff} 
	= \frac{\langle B_r B_{\phi} \rangle}{\Sigma_{\rm g} h_{\rm g} \Omega_K^2} 
	\leq \frac{\langle B \rangle^2}{\Sigma_{\rm g} h_{\rm g} \Omega_K^2} ,
	\label{eq:alpha_eff}
\end{eqnarray}
where $B,\ B_r,\ B_{\phi}$ are the strength, radial component, and azimuthal component of magnetic fields 
 of the solar nebula around the chondrule-forming region, respectively.
Once the value of $\alpha_{\rm eff}$ is given, the scale height of chondrules can be given as \citep{Dubrulle+1995}, 
\begin{eqnarray}
	h_{\rm ch} &=& \frac{H}{\sqrt{1+H^2}} h_{\rm g},
\end{eqnarray}
where $H$ is a quantity derived from the condition that turbulent vertical diffusion ($\alpha_{\rm eff}$) balances out with dust settling toward the midplane,
 which is characterized by $\tau_{\rm stop}$. 
In the actual formula, $H$ can be written as
\begin{eqnarray}
	H &=& \left(\frac{1}{1+\gamma_{\rm turb}}\right)^{1/4} \left(\frac{\alpha_{\rm eff}}{\tau_{\rm stop} \Omega_K } \right)^{1/2} \nonumber \\
	&=& 0.29
		\left(\frac{3}{1+2(\gamma_{\rm turb}/2) }\right)^{1/4} 
		\left( \frac{\langle B \rangle }{\rm 50\ mG } \right) \nonumber\\&&\times
		\left( \frac{\rho_{\rm s}}{3.3 \mbox{ g cm}^{-3} } \right)^{-1/2} 
		\left( \frac{r_{\rm ch}}{\rm 1\ mm} \right)^{-1/2} 
		\left( \frac{r}{\rm 1\ au} \right)^{7/8} ,\nonumber\\
		\label{eq:H}
\end{eqnarray}
where $\gamma_{\rm turn}$ is a quantity related to the nature of turbulence. 
Based on the experimental results obtained from Semarkona ordinary chondrite,
 the typical value of $\langle B \rangle$ is $\langle B \rangle\simeq50$ - 540 mG for the solar nebula \citep{Fu_R+2014}.

\subsection{Growth of a protoplanet}

We use the same model of a protoplanetary growth as the one used in \cite{Hasegawa+2016a} (see their section 2).
We put a protoplanet in a planetesimal swarm. 
The initial mass of the protoplanet is defined as 
\begin{eqnarray}
	M_{\rm ini} &=& 50 m_{\rm pl,min}
		\left( \frac{m_{\rm pl,min}}{10^{23} \mbox{ g}} \right)^{-2/5} \left( \frac{\Sigma_{\rm d}}{10 \mbox{ g cm}^{-2}} \right)^{3/5} \nonumber \\ && \times
		\left( \frac{r}{1 \mbox{ au}} \right)^{6/5} , \label{eq:M_ini}
\end{eqnarray}
where $m_{\rm pl,min}$ is the mass of the smallest planetesimals.
When a protoplanet exceeds this mass, oligarchic growth begins \citep{Ida&Makino1993,Kokubo&Ida1998}.
The accretion rate of a protoplanet ($dM/dt$) is given by 
\begin{eqnarray}
	\frac{dM}{dt} &=& C\pi \Sigma_{\rm d}\frac{2GMR}{ \langle e^2_{\rm pl}\rangle r v_{\rm K} }, 
	\label{eq:dM_dt}
\end{eqnarray}
where $C$ is the accretion acceleration factor, $C=2$, $R$ is the radius of the protoplanet,
 and $\langle e^2_{\rm pl}\rangle^{1/2}$ is the root mean square equilibrium eccentricity of planetesimals. 
The radius of the protoplanet is calculated with $\rho_{\rm pr}=2\mbox{ g cm}^{-3}$, where $\rho_{\rm pr}$ is the material density of the protoplanet.
The equilibrium eccentricity in the oligarchic growth stage is 
\begin{eqnarray}
	\langle e^2_{\rm pl}\rangle^{1/2} 
		&\simeq& 
		5.6\times10^{-2} \left( \frac{m_{\rm pl}}{10^{23} \mbox{ g}} \right) ^{1/15} \left( \frac{\rho_{\rm pl}}{2 \mbox{ g cm}^{-3}} \right) ^{2/15} \nonumber \\ &&\times
		\left( \frac{\rho_{\rm g}}{2\times10^{-9} \mbox{ g cm}^{-3}} \right) ^{-1/5} \left( \frac{r}{1\mbox{ au}} \right)^{-1/5} \nonumber \\&&\times
		\left( \frac{M}{M_{\oplus}} \right)^{1/3}, 
		\label{eq:e_pl}
\end{eqnarray}
where $\rho_{\rm pl}$ is the material density of planetesimals. 
Note that laminar disks are assumed to obtain equation (\ref{eq:e_pl}) \citep{Kokubo&Ida2002}. 
We also assume that a feeding zone of the protoplanet is 10 Hill radius.
The growth of the protoplanet continues until its mass reaches the isolation mass \citep[$M_{\rm iso}$, e.g.,][]{Kokubo&Ida2000},
\begin{eqnarray}
	M_{\rm iso} &=& 0.16M_{\oplus} \left( \frac{\Sigma_{\rm d}}{10\mbox{ g cm}^{-2}} \right)^{3/2} \left( \frac{r}{\mbox{1 au}} \right)^3.
\end{eqnarray}

\subsection{Chondrule formation}\label{sect:chondrule_formation}

In our calculations, we normally adopt the impact jetting model as a chondrule formation model. 
When the impact velocity of planetesimals exceeds 2.5 km s$^{-1}$, chondrules are formed \citep{Johnson+2015, Wakita+2016L, Wakita+2016b}. 
The impact velocity ($v_{\rm imp}$) is given by $v_{\rm imp}=\sqrt{v_{\rm esc}^2+(\langle e_{\rm pl}^2 \rangle^{1/2}v_K)^2}$, where $v_{\rm esc}$ is the escape velocity.
We consider protoplanet-planetesimal collisions as chondrule forming impacts.
This is because planetesimal-planetesimal collisions are much less effective in generating chondrules than protoplanet-planetesimal ones \citep{Hasegawa+2016a}.
In this situation, the mass of chondrules produced during $dt$ becomes $F_{\rm ch} dM$,
 where $F_{\rm ch}$ is the mass fraction of chondrules generated by a jetting collision.
When we consider protoplanet-planetesimal collisions and a threshold velocity for chondrule forming impacts as 2.5 km s$^{-1}$,
 $F_{\rm ch}\simeq 0.01$ \citep{Johnson+2015, Wakita+2016L, Wakita+2016b}, which is adopted in our calculations.
When the mass of the protoplanet reaches the isolation mass, the mass of the cumulative formed chondrules is $\simeq0.01M_{\rm iso}$.

The timescale of protoplanet growth ($\tau_{\rm pr}$) is 
\begin{eqnarray}
	\tau_{\rm pr} &=& f_{\tau} \frac{M}{dM/dt} \nonumber \\
	&=& 2.7 \times 10^5 \times f_{\tau}
	f_d^{-7/5}
	 \left( \frac{m_{\rm pl,min}}{\rm 10^{23}\ g} \right)^{2/15} \nonumber \\ &&\times 
	 \left( \frac{\rho_{\rm pl}}{\rm 2\ g\ cm^{-3}} \right)^{4/15}
	 \left( \frac{r }{\rm 1 \ au } \right)^{27/10} 
	 \left( \frac{M}{0.1M_{\oplus}} \right)^{1/3} \nonumber \\ &&\times 
	 \left( \frac{\rho_{\rm pr}}{\rm 2\ g\ cm^{-3}} \right)^{1/3} \ {\rm yr} ,
	 \label{eq:tau_pr}
\end{eqnarray}
where $f_{\tau}$ is a correcting factor, $f_{\tau}=3$ \citep{Hasegawa+2016a}.
In the impact jetting model, the timescale of chondrule formation is between when mass of protoplanet reaches $M=M_{\rm esc}\simeq0.018M_{\oplus}$,
 which is the mass that the escape velocity becomes equal to 2.5 km s$^{-1}$, and $M=M_{\rm iso}$.

\begin{figure}[ht]
	\plotone{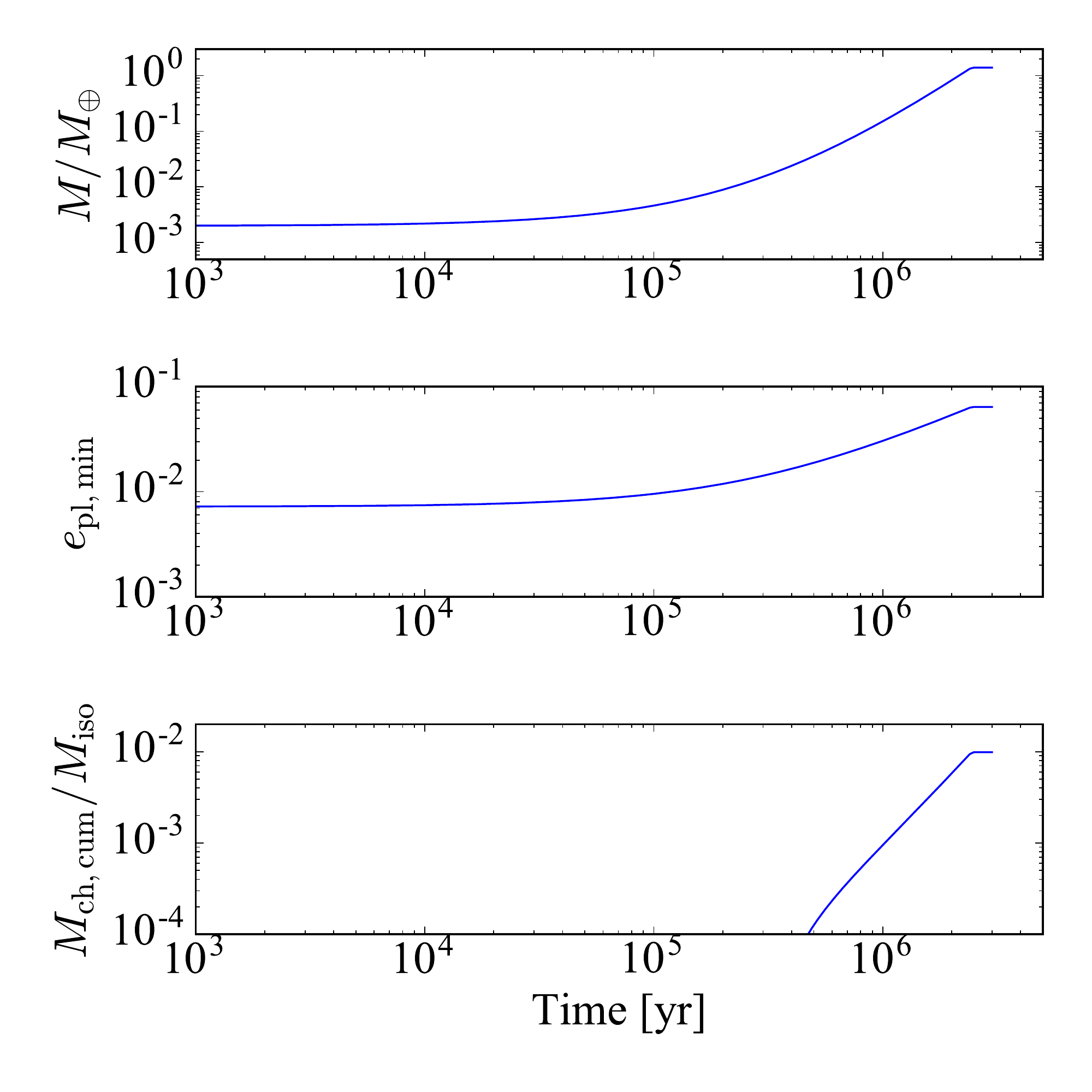}
	\caption{Time evolution of $M$ (top), $e_{\rm pl,min}$ (middle), and the mass of cumulative formed chondrules ($M_{\rm ch,cum}$) (bottom) in our fiducial model.}
	\label{Fig:evolution_sample}
\end{figure}

Figure \ref{Fig:evolution_sample} shows time evolutions of the mass of the protoplanet ($M$), the eccentricity of the smallest planetesimals ($e_{\rm pl, min}$),
 and the mass of cumulative formed chondrules ($M_{\rm ch,cum}$) in our fiducial model.
Since the eccentricities of planetesimals follow the Rayleigh distribution \citep{Ida&Makino1993}, $e_{\rm pl}\simeq\langle e_{\rm pl}^2\rangle^{1/2}$. 
The collision velocity exceeds 2.5 km s$^{-1}$ at $3.3\times10^5$ yr. 
The protoplanet reaches the isolation mass, which is $1.4M_{\oplus}$ at a time of $t_{\rm iso}=2.4\times10^6$ yr.
Chondrules are formed during a span of $2\times10^6$ yr, which is consistent with the formation timescale of chondrules suggested from chondrites.
The mass of cumulative formed chondrules is $0.99\times10^{-2}M_{\rm iso}\simeq F_{\rm ch}M_{\rm iso}$. 

We also consider two different models for chondrule formation. 
In these models, the production rate of chondrules is different from that of the impact jetting model. 
In the first model, we assume that chondrules are formed with a constant rate during $M_{\rm esc} \le M < M_{\rm iso}$. 
This model is hereafter referred to as the constant production rate model. 
In the second model, it is assumed that the production rate decreases linearly with time. 
This model is hereafter called as the decreasing production rate model. 
Since the production rate of the impact jetting model increases with time ($F_{\rm ch} dM$), we can examine all the three distinct models for chondrule formation. 
Note that the total mass of chondrules formed in all the models is about $0.01M_{\rm iso}$;
 in the constant production rate model, $\simeq F_{\rm ch}M_{\rm iso}/2\times10^6 \mbox{ yr}\simeq4\times10^{19}$ g of chondrules are formed per year,
 while in the decreasing production rate model, the mass of chondrules formed at $M=M_{\rm esc}$ is $\simeq 7\times10^{19}$ g,
 which is ten times larger than that at $M=M_{\rm iso}$.

\subsection{Chondrule accretion}\label{sect:ch_acc}

In the following, we describe how a protoplanet and planetesimals accrete chondrules.
Our estimation is based on \cite{LJ2012}. 
In order to explicitly compare the accretion efficiencies of chondrules by a protoplanet with that by planetesimals,
 we assume that these massive objects are exposed to the same amount of chondrules. 
In other words, we estimate the accretion timescale of chondrules by a protoplanet and by planetesimals, independently. 
Chondrule masses accreted by a protoplanet and by planetesimals are derived from these timescales.

\subsubsection{Protoplanet}\label{sect:chondrule_acc_pr}

The relative velocity ($\Delta v$) between an accreting body and chondrules is important to estimate chondrule accretion.
The relative velocity is caused by the eccentricity of the body, gas drag, and Keplerian shear.
In our simulations, $\Delta v$ between the protoplanet and chondrules is written as $\eta v_K$.
The eccentricity of the protoplanet is $\sim\sqrt{m_{\rm pl,min}/M }e_{\rm pl,min}$ by energy equipartition. 
In our parameter range, $\eta$ is larger than the eccentricity of the protoplanet and Keplerian shear.
If we consider larger pebbles or a larger protoplanet, $\Delta v$ is determined by Keplerian shear as in the case of the estimation done by \cite{LJ2012}.

Disk turbulence excites the eccentricity of a protoplanet \citep{Ida+2008}. 
However, the turbulence is weak ($4\times10^{-5}\leq \alpha_{\rm eff}\leq 5\times10^{-3}$) when $50\mbox{ mG}\leq \langle B\rangle\leq 540$ mG in the solar nebula \citep{Fu_R+2014}. 
In addition, a longer time is needed for a protoplanet to experience the eccentricity pump-up by disk turbulence
 than to undergo the eccentricity damping by dynamical friction from planetesimals. 
We do not consider the effect of turbulence on the protoplanet, and hence $\Delta v=\eta v_K$. 

There are two modes when a protoplanet accretes chondrules \citep{LJ2012}: the drift accretion mode, and the Hill accretion one. 
These two modes are divided by a transition mass ($M_t$). 
Comparing the Bondi radius $r_{\rm B} = GM/\Delta v^2$ with the Hill radius $r_{\rm H} = (M/3M_{\odot})^{1/3} r$, we can get the transition mass ($M_{\rm t}$), 
\begin{eqnarray}
	M_{\rm t} &=& \frac{\Delta v^3}{\sqrt{3}G\Omega_K} 
	= 1.1\times 10^{-3} \left( \frac{r}{1 \mbox{ au} } \right)^{3/2} M_{\oplus} ,
\end{eqnarray}
which is the mass that $r_{\rm B}=r_{\rm H}$.
Since $M_{\rm esc}>M_{\rm t}$, the protoplanet is in the Hill accretion mode \citep{LJ2012}.
Given that chondrule are well coupled with gas (see equation \ref{eq:tau_stop}), the accretion radius ($r_{\rm acc}$) of chondrules by a protoplanet in the Hill mode is determined as what follows;
 chondrule accretion by a protoplanet can be achieved when the timescale 
 that the gravitational pull arising from a protoplanet can affect chondrules' orbits becomes comparable to the stopping time of chondrules:
 \footnote{In \cite{Ormel&Klahr2010}, this accretion process is named as settling, since  chondrules reside in the strong coupling regime.}
\begin{eqnarray}
	\frac{\Delta v}{GM/r_{acc}^2} &=& \tau_{\rm stop} \nonumber \\
	\ \Leftrightarrow \ 
	r_{acc} 
		&=& \sqrt{ \tau_{\rm stop} \frac{GM}{\eta r\Omega_K} }\nonumber \\
		&=& 7.2\times10^{-2} \ f_d^{-1/2}
		\left( \frac{r_{\rm ch}}{\mbox{1 mm}} \right)^{1/2} \nonumber \\&&\times
		\left( \frac{\rho_s}{\mbox{3.3 g cm}^{-3}} \right)^{1/2} 
		\left( \frac{r }{\mbox{1 au}} \right)^{1/2}
		\left( \frac{M}{M_{\oplus}} \right)^{1/6} \ r_{\rm H}.\nonumber \\
	\label{eq:r_acc_pr}
\end{eqnarray}
Substituting $r_{\rm H}=1.0\times10^{-2} \left( M/M_{\oplus} \right)^{1/3}r$, $r_{\rm acc}=7.2\times10^{-4} (M/M_{\oplus})^{1/2} (r/1\mbox{ au})^{3/2}$ au.
The chondrule accreting rate by the protoplanet (${\dot M_{\rm acc, pr}}$) is
 ${\dot M_{\rm acc, pr}} = \pi \rho_{\rm ch}r_{acc}^2\Delta v$, where $\rho_{\rm ch}$ is the spatial density of chondrules. 
The density of chondrules can be given as 
 $\rho_{\rm ch}=M_{\rm ch}/( 2\pi^{3/2}r\Delta r h_{\rm ch} )$, where $\Delta r$ is the orbital width that chondrules are distributed in, and we give $\Delta r=h_{\rm ch}$.
Note that a specific choice of $\Delta r$ does not affect our conclusions,
 because the accretion timescales of chondrules both by a protoplanet and by planetesimals have the same dependence on $\rho_{\rm ch}$ (see below).

Now, we derive the accretion rate (${\dot M_{\rm acc, pr}}$) and the timescale ($\tau_{\rm acc, pr}$) of chondrules accreted by a protoplanet. 
Considering the protoplanet at 2 au and $H=0.53$, ${\dot M_{\rm acc,pr}}$ becomes
\begin{eqnarray}
	{\dot M_{\rm acc, pr} }
	&=& \pi \rho_{\rm ch}r_{acc}^2\Delta v\nonumber \\
	&\simeq& \pi \left( \frac{ M_{\rm ch} }{ 2\pi^{3/2}r h_{\rm ch}^2 }\right) r_{acc}^2\eta v_K \nonumber \\
	&=& 3.1\times 10^{-7} \left(\frac{f_d}{3}\right)^{-1} 
		\left( \frac{ H^2/(1+H^2) }{0.25} \right)^{-1} \left( \frac{r }{\mbox{2 au}} \right) \nonumber \\ &&\times
		\left( \frac{r_{\rm ch}}{\mbox{1 mm}} \right) \left( \frac{\rho_s}{\mbox{3.3 g cm}^{-3}} \right)
		\left( \frac{M}{M_{\rm esc}} \right) T_K^{-1} M_{ch}. \nonumber \\
		\label{eq:dM_acc_pr}
\end{eqnarray}
The timescale of chondrule accretion by the protoplanet is determined by $\tau_{\rm acc, pr}\equiv M_{\rm ch}/{\dot M_{\rm acc, pr}}$, 
\begin{eqnarray}
	\tau_{\rm acc,pr} 
	&=& 0.91\times 10^{7} 
		\left(\frac{f_d}{3}\right)
		\left( \frac{ H^2/(1+H^2) }{0.25} \right) \left( \frac{r }{\mbox{2 au}} \right)^{1/2} \nonumber \\&&\times
		\left( \frac{r_{\rm ch}}{\mbox{1 mm}} \right)^{-1} \left( \frac{\rho_s}{\mbox{3.3 g cm}^{-3}} \right)^{-1} \left( \frac{M}{M_{\rm esc}} \right)^{-1} 
		\mbox{ yr}. \nonumber \\
		\label{eq:tau_acc_pr}
\end{eqnarray}
Since the chondrule accretion radius becomes larger with increasing $M$ (see equation \ref{eq:r_acc_pr}), $\tau_{\rm acc,pr}$ decreases with increasing $M$.

\subsubsection{Planetesimals}\label{sect:chondrule_acc_pl}

Next, we consider chondrule accretion by planetesimals. 
While we follow a basic formalism that has been developed by \cite{LJ2012} and \cite{Ormel&Klahr2010},
 the picture of chondrule accretion by planetesimals in our estimation is different from theirs. 
In the oligarchic growth, random velocities and numbers of planetesimals are changed according to the mass growth of the protoplanet.
These largely affect the chondrule accretion rate of planetesimals.

When the mass of planetesimals exceeds $M_{\rm t}$, 
 the accretion radius of chondrules is described in the same way as that of a protoplanet (see equations \ref{eq:r_acc_pr} and \ref{eq:tau_acc_pr}). 
In the following, we consider planetesimals that have smaller masses than $M_{\rm t}$, i.e., in the drift accretion mode \citep{LJ2012}. 
In this mode, chondrule accretion radius is determined according to $\tau_{\rm B}/\tau_{\rm stop}$, where $\tau_{\rm B}=r_{\rm B}/\Delta v$ \citep{LJ2012}.
When $1<\tau_{\rm B}/\tau_{\rm stop}$, chondrules are strongly affected by gas drag, and planetesimals can not accrete chondrules in whole $r_{\rm B}$. 
This case corresponds to the settling regime in \cite{Ormel&Klahr2010} (also see Section \ref{sect:chondrule_acc_pr}). 
For this case, the accretion radius is determined by the balance between gravitational pull from a planetesimal and gas-drag acting on chondrules.
The accretion radius increases up to $r_{\rm B}$ as $\tau_{\rm B} / \tau_{\rm stop}$ decreases. 
Since we consider chondrules (that is, a constant value of $\tau_{\rm stop}$), $\tau_{\rm B} / \tau_{\rm stop}$ decreases as $m_{\rm pl}$ becomes smaller or $\Delta v$ becomes larger. 
For the case that $r_{\rm acc} = r_{\rm B}$, chondrule accretion becomes the most efficient in the sense that all the chondrules in the Bondi radius will spiral towards planetesimals. 
This arises because chondrules experience less gas-drag as their orbit is deflected by planetesimals. 
This settling regime continues until $\tau_{\rm B} / \tau_{\rm stop} \simeq 0.25$ at which gravitational focusing of a planetesimal regulates the dynamics of chondrules. 
For this case, the accretion radius is given by the gravitational focusing. 
This case is called the hyperbolic regime, and planetesimals are in this regime when $\tau_{\rm B}/\tau_{\rm stop}<0.25$ \citep{Ormel&Klahr2010}.
In the hyperbolic regime, the orbit of a pebble is determined only by the gravitation interaction of a large body,
 while in the settling regime, that is affected both by gas-drag and by the gravitational interaction, which is called pebble accretion in \cite{LJ2012}.

\begin{figure}[ht]
	\plotone{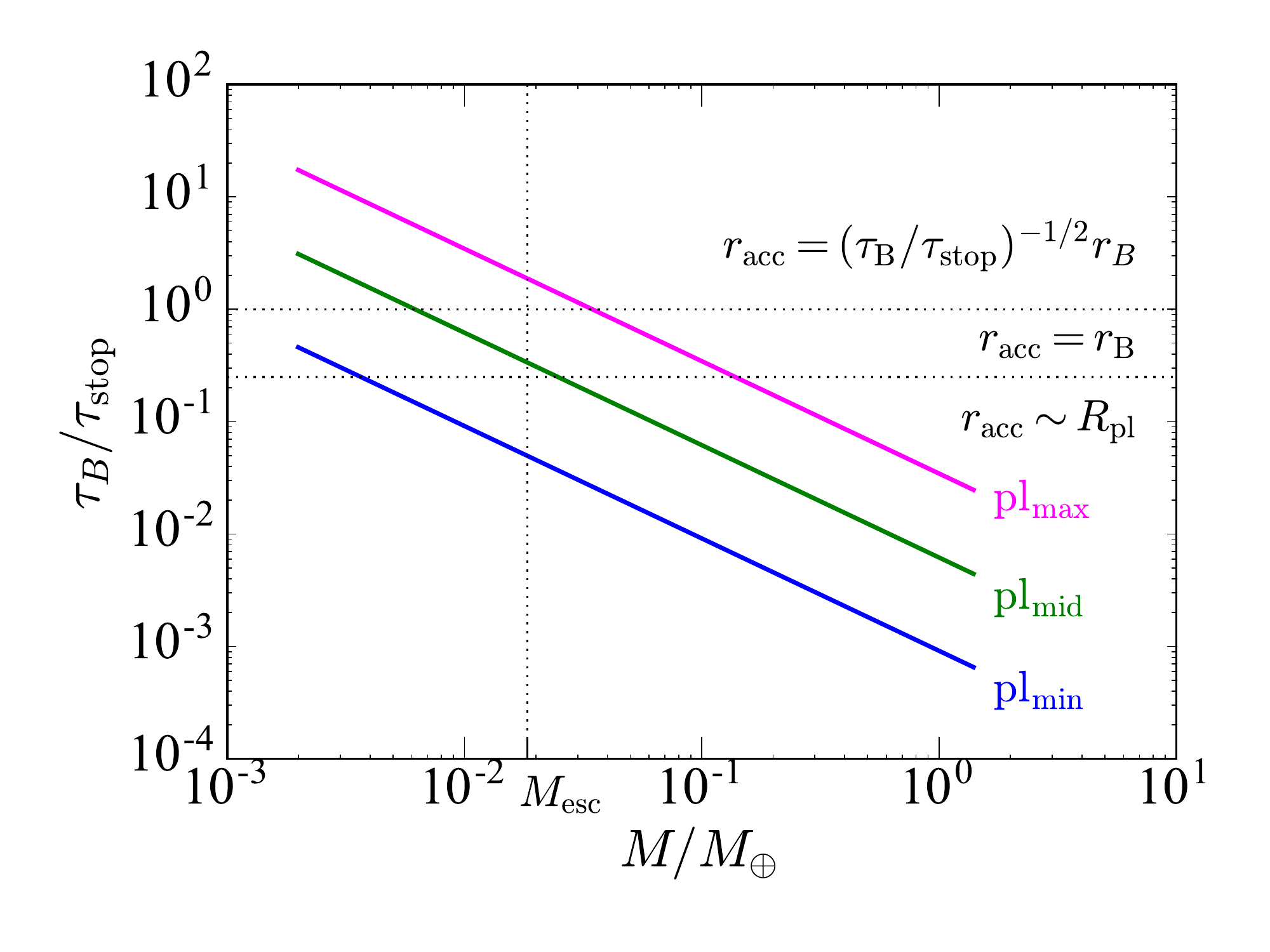}
	\caption{
		The evolution of $\tau_{\rm B}/\tau_{\rm stop}$ of planetesimals is shown as a function of $M/M_{\oplus}$ in our fiducial model. 
		The three lines are $\tau_{\rm B}/\tau_{\rm stop}$ of the smallest planetesimals (pl$_{\rm min}$, the blue line), median mass planetesimals (pl$_{\rm mid}$, the green line),
		 and the largest planetesimals (pl$_{\rm max}$, the purple line) in ascending order. 
		These lines are plotted between $M_{\rm ini}$ and $M_{\rm iso}$. 
		When $M\geq M_{\rm esc}$, chondrules are formed (the vertical dotted line).
		The two dotted horizontal lines identify the three accretion cases at which the accretion radii are different from each other.
	}
	\label{Fig:tauB_tau_stop}
\end{figure}

The relative velocity between planetesimals and chondrules is $\Delta v=e_{\rm pl}v_K$ for all the three cases
 ($r_{acc} = (\tau_{\rm B}/\tau_{\rm stop})^{-1/2} r_B, r_{\rm acc} = r_{\rm B},$ and $r_{\rm acc} \sim R_{\rm pl}$). 
This is because $e_{\rm pl}$ is larger than $\eta$ and Kepler shear.
Since eccentricities of planetesimals increase according to $M$ (equation \ref{eq:e_pl}), $\tau_{\rm B}/\tau_{\rm stop}$ is changed as the protoplanet mass ($M$) increases,
\begin{eqnarray}
	\frac{\tau_{\rm B}}{\tau_{\rm stop}}
		&=& 2.7\times10^{-2} f_d \left( \frac{m_{pl}}{10^{23} \mbox{ g}} \right) ^{4/5} \left( \frac{\rho_{pl}}{2 \mbox{ g cm}^{-3}} \right) ^{-2/5} \nonumber \\&&\times
		\left( \frac{\rho_{\rm g}}{2\times10^{-9} \mbox{ g cm}^{-3}} \right) ^{3/5} \left( \frac{r}{2\mbox{ au}} \right)^{-9/10} 
		\nonumber \\&&\times \left( \frac{M}{M_{\rm esc}} \right)^{-1} 
		\left( \frac{r_{\rm ch} }{\mbox{1 mm}} \right)^{-1} \left( \frac{\rho_s}{\mbox{3.3 g cm}^{-3}} \right)^{-1} . \nonumber \\
		\label{eq:tauB_tau_stop}
\end{eqnarray}
As explicitly seen in equation (\ref{eq:tauB_tau_stop}), $\Delta v$ increases and $\tau_{\rm B}$ becomes smaller following the mass growth of protoplanets. 
Figure \ref{Fig:tauB_tau_stop} shows $\tau_{\rm B}/\tau_{\rm stop}$ as a function of $M$ in our fiducial model.
We set 20 bins between $m_{\rm pl,min}$ and $M_{\rm ini}$ (see equation \ref{eq:M_ini}). 
During chondrule formation ($M>M_{\rm esc}$), $\tau_{\rm B}/\tau_{\rm stop}$ of the smallest planetesimals (pl$_{\rm min}$) is always smaller than 0.25 (that is, the hyperbolic regime) . 
The median mass planetesimals (pl$_{\rm mid}$), which have $\sqrt{m_{\rm pl,min}M_{\rm ini}}$ mass, also spend most of the span of chondrule formation in the hyperbolic regime. 
In this figure, only the largest planetesimals, which have $m_{\rm pl,min}^{1/20}M_{\rm ini}^{19/20}$ mass, accrete chondrules via the pebble accretion. 

In our simulations, the size distribution of planetesimals is taken into account when the accretion timescale is estimated. 
The number of planetesimals is given by the power law, $n_{\rm pl}= f_n (m_{\rm pl}/M_{\rm ini})^{-2}$,
 where $n_{\rm pl}$ is the number of planetesimals in a bin \citep{Kokubo&Ida2000, Morishima+2008}.
To keep the total mass of planetesimals ($\sum m_{\rm pl} n_{\rm pl}$) constant for all the simulations, $n_{\rm pl}$ is multiplied by a factor $f_n$. 
This factor is approximately proportional to $m_{\rm pl,min}^{-1/5} \Sigma_{\rm d}^{-6/5}r^{-12/5}$.
In our fiducial model, $f_n=1$, and the total mass of planetesimals always corresponds to the one in our fiducial model if $f_d=3$. 
While the size distribution of planetesimals is included in our estimate,
 it is reasonable to assume that planetesimals in each mass bin accrete chondrules from their whole $r_{\rm acc}$.
This is because planetesimals' cross-sections of accretion are much smaller than $2\pi r\Delta r$.
Then the mass accretion rate of chondrules by planetesimals in each bin is computed as the summation of that by each planetesimal.
Since the protoplanet's cross-section of accretion is also much smaller than $2\pi r\Delta r$, 
 we assume that a protoplanet and planetesimals do not compete in accreting chondrules. 
Also, to accurately estimate the accretion efficiency of chondrules only by planetesimals,
 the reduction of $n_{\rm pl}$ due to the protoplanet growth is neglected in our simulations. 
In other words, both a protoplanet and planetesimals are exposed to the same amount of chondrules.
This assumption may end up with that the total mass of chondrules accreted by planetesimals may be overestimated. 
Nonetheless, our estimate is useful in the sense that once the total amount of chondrules accreted by single planetesimals is obtained,
 we can readily calculate how much of chondrules are eventually accreted by planetesimals in each mass bin.

We now derive the accretion radius ($r_{\rm acc}$) of chondrules by planetesimals and its timescale ($\tau_{\rm acc, pl}$). 
At first, we consider $1<\tau_{\rm B}/\tau_{\rm stop}$. 
In this case, the chondrule accretion radius is (as done in Section \ref{sect:chondrule_acc_pr}) 
\begin{eqnarray}
	\frac{\Delta v}{ G m_{\rm pl}/r_{\rm acc}^2 } &=& \tau_{\rm stop} \
	\Leftrightarrow \ r_{\rm acc} = \left( \frac{\tau_{\rm B}}{\tau_{\rm stop}} \right)^{-1/2} r_{\rm B},
	\label{eq:r_acc_pl_0}
\end{eqnarray}
since the orbits of chondrules are both affected by the gas drag and planetesimal gravity. 
Here, the Bondi radius of a planetesimal is $r_{\rm B}= Gm_{\rm pl}/\Delta v^2$. 
In this situation, the chondrule accreting rate by planetesimals (${\dot M_{\rm acc, pl}}$) is
 ${\dot M_{\rm acc, pl}} = n_{\rm pl} \pi \rho_{\rm ch} \left[(\tau_{\rm B}/\tau_{\rm stop})^{-1/2} r_{\rm B}\right]^2 e_{\rm pl} v_K$.
The timescale of chondrule accretion by planetesimals is
\begin{eqnarray}
	\tau_{\rm acc, pl} 
		&\equiv& M_{\rm ch}/{\dot M_{\rm acc, pl}} \nonumber \\
		&=& 1.7\times10^{7} f_n^{-1}\left(\frac{f_d}{3}\right) \left( \frac{m_{\rm pl}/M_{\rm ini}}{1/120} \right)^2
			\left( \frac{H^2/(1+H^2) }{0.25} \right) \nonumber \\&& \times
			\left( \frac{m_{\rm pl}}{10^{23} \mbox{ g}} \right) ^{-1} 
			\left( \frac{\rho_{\rm g}}{2\times10^{-9} \mbox{ g cm}^{-3}} \right)^{8/5}
			\left( \frac{r}{2\mbox{ au}} \right)^{21/10} \nonumber \\&& \times
			\left( \frac{r_{\rm ch}}{\mbox{1 mm}} \right)^{-1} \left( \frac{\rho_s}{\mbox{3.3 g cm}^{-3}} \right)^{-1} \mbox{ yr}.
	\label{eq:tau_acc_pl_0}
\end{eqnarray}
When $0.25<\tau_{\rm B}/\tau_{\rm stop}<1$, planetesimals accrete chondrules from the whole Bondi radius, $r_{\rm acc}=r_{\rm B}$ \citep{Ormel&Klahr2010}
In this case, $\tau_{\rm acc, pl}$ becomes
\begin{eqnarray}
	\tau_{\rm acc, pl} 
		&=& 2.1
			\times10^{8}f_n^{-1}\left( \frac{m_{\rm pl}/M_{\rm ini}}{1/120} \right)^2
			\left( \frac{H^2/(1+H^2) }{0.25} \right) \nonumber \\&& \times
			\left( \frac{m_{\rm pl}}{10^{23} \mbox{ g}} \right) ^{-27/15} 
			\left( \frac{\rho_{\rm pl}}{2 \mbox{ g cm}^{-3}} \right) ^{2/3} 
			\left( \frac{r}{2\mbox{ au}} \right)^{3}\nonumber \\&& \times
			\left( \frac{\rho_{\rm g}}{2\times10^{-9} \mbox{ g cm}^{-3}} \right) 
			\left( \frac{M}{M_{\rm esc}} \right) \mbox{ yr}.
	\label{eq:tau_acc_pl_1}
\end{eqnarray}
When $\tau_{\rm B}/\tau_{\rm stop}<0.25$, the chondrule accretion is in hyperbolic regime,
 and the gravitational scattering plays the dominant role in accreting chondrules. 
Planetesimals can accrete chondrules only from the gravitationally-enhanced cross section, $R_{\rm pl}\sqrt{1+(v_{\rm esc}/\Delta v)^2}$,
 where $R_{\rm pl}$ is the radius of planetesimals \citep{Ormel&Klahr2010}. 
The timescale of the chondrule accretion is 
\begin{eqnarray}
	\tau_{\rm acc, pl} 
		&=& 1.2\times10^{7}f_n^{-1}\left( \frac{m_{\rm pl}/M_{\rm ini}}{1/120} \right)^2
			\left( \frac{H^2/(1+H^2) }{0.25} \right) \nonumber \\&& \times
			\left( \frac{m_{\rm pl}}{10^{23} \mbox{ g}} \right) ^{-11/15} 
			\left( \frac{\rho_{\rm pl}}{2 \mbox{ g cm}^{-3}} \right) ^{8/15}
			\left( \frac{r}{2\mbox{ au}} \right)^{21/5} \nonumber \\&& \times
			\left( \frac{\rho_{\rm g}}{2\times10^{-9} \mbox{ g cm}^{-3}} \right)^{1/5} 
			\left( \frac{M}{M_{\rm esc}} \right)^{-1/3} \nonumber \\&& \times
			\left( 1+\left(\frac{v_{\rm esc}}{e_{\rm pl} v_K}\right)^2 \right)^{-1}
			\mbox{ yr}.
	\label{eq:tau_acc_pl_2}
\end{eqnarray}

Orbital inclinations can also affect chondrule accretion \citep{Levison+2015}. 
This quantity comes into play in our model, because planetesimals and a protoplanet coexist in the system. 
When the inclinations of the planetesimals are larger than $h_{\rm ch}/r$, the planetesimals can not accrete chondrules in whole their orbits.
We calibrate the effect of the inclination by computing the ratio of the orbital period to a time interval
 during which planetesimals reside within the height of $h_{\rm ch}$ from the midplane.
The inclinations of planetesimals are given by $i_{\rm pl}=e_{\rm pl}/2$. 
Using Hill's equations \citep{Nakazawa&Ida1988}, this ratio can be described as 
\begin{eqnarray}
	f_{i_{\rm pl}} &=& \frac{ \frac{4}{\Omega_K} \mbox{ asin}{\left( \frac{ h_{ch} }{ i_{pl}r }\right)} }{T_K} = \frac{2}{\pi} \mbox{ asin}{\left( \frac{ h_{ch} }{ i_{pl}r }\right)}, 
	\label{eq:tau_acc_f_i}
\end{eqnarray}
by which ${\dot M_{\rm acc}}$ is multiplied, when $ri_{\rm pl}>h_{\rm ch}$.
The derivation of $f_{i_{\rm pl}}$ is summarized in Appendix. 

\subsubsection{The resultant timescale of accreting chondrules}\label{sect:timescale_chondrule_acc}

\begin{figure}[ht]
	\plotone{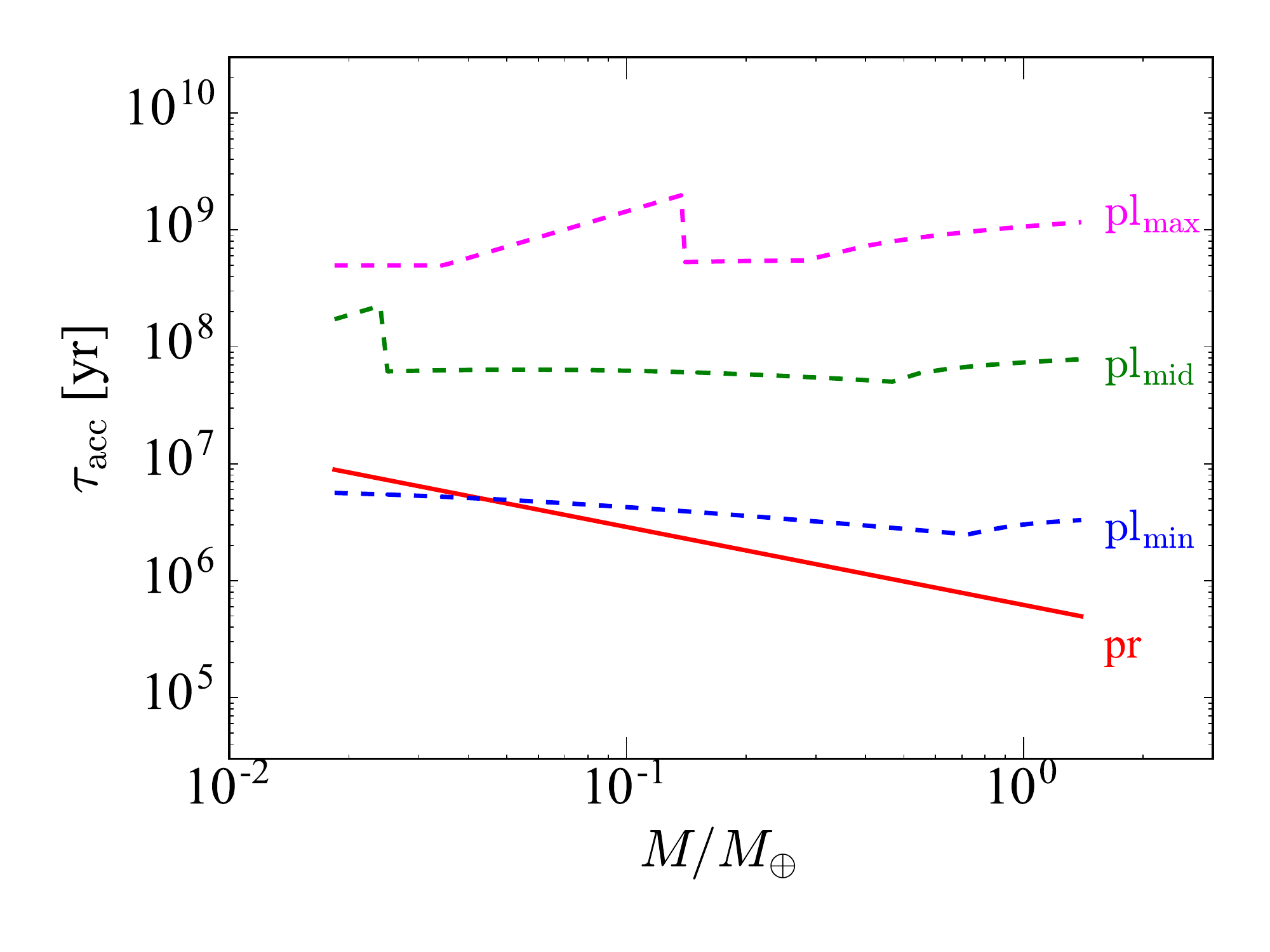}
	\caption{
	The timescales of chondrule accretion are shown as a function of $M$ in our fiducial model.
	The solid line with index pr is that of the protoplanet ($\tau_{\rm acc,pr}$, equation \ref{eq:tau_acc_pr}),
	 and the dashed lines are those of planetesimals ($\tau_{\rm acc,pl}$) in each mass range 
	 (the purple dashed line is that of pl$_{\rm max}$, the green dashed line is that of pl$_{\rm mid}$, and the blue dashed line is that of pl$_{\rm min}$). 
	These lines are plotted between $M_{\rm esc}$ and $M_{\rm iso}$.
	According to $\tau_{\rm B}/\tau_{\rm stop}$, $\tau_{\rm acc,pl}$ is determined by equations (\ref{eq:tau_acc_pl_0}), (\ref{eq:tau_acc_pl_1}), and (\ref{eq:tau_acc_pl_2}).
	When $ri_{\rm pl}\geq h_{\rm ch}$, where the planetesimals can not accrete chondrules in whole their orbits,
	 $\tau_{\rm B}/\tau_{\rm stop}$ increases with $M$ (equation \ref{eq:tau_acc_f_i}).
	}
	\label{Fig:M_vs_tau_acc}
\end{figure}

The timescales of chondrule accretion in the fiducial model are shown in Figure \ref{Fig:M_vs_tau_acc}. 
These timescales by a protoplanet and by planetesimals in each mass range are plotted as a function of $M$. 
The timescale by the protoplanet decreases with increasing $M$ (see the red solid line).
This is because $r_{\rm acc}$ increases with increasing $M$ (equations \ref{eq:r_acc_pr} and \ref{eq:tau_acc_pr}). 
In the case of pl$_{\rm min}$ planetesimals (see the blue dashed line),
 planetesimals are in the hyperbolic regime, and $\tau_{\rm acc,pl}$ is given by equation (\ref{eq:tau_acc_pl_2}).
In this regime, $\tau_{\rm acc,pl}$ depends on $M$ only through $e_{\rm pl}$, which increases with increasing $M$ (equation \ref{eq:e_pl}).
As a result, the planetesimals can encounter more chondrules as a protoplanet becomes more massive.
This is why $\tau_{\rm acc,pl}$ decreases gradually with increasing $M$ when $M<0.7M_{\oplus}$. 
When $M>0.7M_{\oplus}$, the inclination of pl$_{\rm min}$ planetesimals becomes larger than $h_{\rm ch}/r$. 
For this case, the planetesimals have less chance to accrete chondrules, simply because the planetesimals can stay in the chondrule sea for a shorter time. 
Consequently, $\tau_{\rm acc}$ becomes longer.
The effect of the inclination (equation \ref{eq:tau_acc_f_i}) increases $\tau_{\rm acc,pl}$ as increasing $M$. 
For pl$_{\rm mid}$ planetesimals (see the green line), two similar features are seen in the behavior of $\tau_{\rm acc}$, compared with the pl$_{\rm min}$ planetesimal case;
 the first one is that the accretion timescale decreases slowly with increasing $M$ when $0.025 < M/M_{\oplus}< 0.47$. 
This is again because the planetesimals are in the hyperbolic regime. 
The other feature is that $\tau_{\rm acc}$ increases with $M$, which is caused by the inclination effect.
Since the inclination of pl$_{\rm mid}$ planetesimals grows faster than that of pl$_{\rm min}$ planetesimals,
 the effect of $f_{i_{\rm pl}}$ becomes important when the protoplanet reaches $0.4M_{\oplus}$.
There is another noticeable feature for the case of pl$_{\rm mid}$; the accretion timescale jumps at $M=0.02M_{\oplus}$. 
This jump is caused by discontinuous change of $r_{\rm acc}$ between the settling regime and the hyperbolic regime
 which occurs at $\tau_{\rm B} / \tau_{\rm stop}$ at 0.25 (see Figure \ref{Fig:tauB_tau_stop}).
The same jump is also seen in $\tau_{\rm acc,pl}$ of pl$_{\rm max}$ planetesimals as well (see the purple dashed line).
For pl$_{\rm max}$ planetesimals, $\tau_{\rm acc,pl}$ is constant when $M<0.03M_{\oplus}$ (Equation \ref{eq:tau_acc_pl_0}). 
In this mass range, $1<\tau_{\rm B}/\tau_{\rm stop}$, and hence the accretion radius is smaller than the Bondi radius (see equation \ref{eq:r_acc_pl_0}). 
Since the relative velocity is determined by $e_{\rm pl} v_{K}$, $r_{\rm acc} \propto e_{\rm pl}^{-1/2}$, it indicates that $r_{\rm acc}$ shirks with increasing $M$. 
At the same time, however, $\tau_{\rm acc, pl} \propto r_{\rm acc,pl}^2 \Delta v$. 
As a result, the accretion timescale in this case does not depend on $M$. 
After the protoplanet has larger masses than $0.03M_{\oplus}$, $\tau_{\rm acc,pl}$ increases with increasing $M$ (equation \ref{eq:tau_acc_pl_1}),
 since $r_{\rm acc}=r_{\rm B}$ in this regime and $e_{\rm pl}$ dependence on $\tau_{\rm acc,pl}$ is not canceled out anymore.
When $M>0.14M_{\oplus}$, $\tau_{\rm acc,pl}$ evolves according to equation (\ref{eq:tau_acc_pl_2}), that is the hyperbolic regime.
Figure \ref{Fig:M_vs_tau_acc} shows that $\tau_{\rm acc,pr}$ is shorter than any $\tau_{\rm acc,pl}$ when $M>0.04M_{\oplus}$. 
This suggests that most chondrules would be accreted by a protoplanet. 
For planetesimals, $\tau_{\rm acc,pl}$ of pl$_{\rm min}$ planetesimals is the smallest. 
While the timescale of chondrule accretion by a single planetesimal becomes shorter with increasing $m_{\rm pl}$,
 $\tau_{\rm acc,pl}$ of planetesimals in each mass bin becomes longer with increasing $m_{\rm pl}$.
This is simply because the number of planetesimals is taken into account when computing $\tau_{\rm acc, pl}$.

In some of the following simulations, the effect of disk turbulence on chondrule accretion by planetesimals will be examined
 by multiplying $\tau_{\rm acc, pl}$ by a factor of $f_{\rm acc}^{-1}$. 
This is because a number of effects triggered by disk turbulence have been proposed. 
These include that chondrules can be concentrated by weak turbulence \citep[e.g., ][]{Cuzzi+2001}, that eccentricities of planetesimals are excited by turbulence \citep{Ida+2008},
 and that the collision probability between planetesimals and chondrules is changed by turbulence \citep{Guillot+2014}. 
In this paper, we take into account only a turbulent effect that can change the collisional probability. 
This can be done by changing the value of $f_{\rm acc}$.
The concentration process of chondrules by turbulence in the oligarchic growth would be affected by protoplanets. 
While random torque arising from disk turbulence can pump up planetesimals' eccentricities,
 the eccentricity excitation by a protoplanet is likely to be more important in our configuration \citep{Hasegawa+2016b}. 
Thus, the concentration of chondrules and eccentricity excitation by turbulence are not included in our simulations.
Note that the estimation of $h_{\rm ch}$ includes the turbulent effect \citep{Dubrulle+1995}.

\section{Chondrule formation and accretion}\label{sect:chondrule_acc}

\begin{deluxetable*}{lccccccp{18em}}
	\tablenum{2}
	\tablecaption{Summary of Simulations}\label{table:models}
	\tablehead{
	Section					&	$m_{\rm pl,min}$		&	$f_{\rm d}$	&	$r$		&	$\tau_{\rm g}$	&	$f_{\rm acc}$	&	$F_{\rm ch}$	&	Chondrule formation model
	}
	\startdata
	\S\ref{sect:fiducial} (fiducial)	&	$10^{23}$ g			&	3			&	2 au		&	$\infty$			&	1		&	0.01			&	Impact jetting	\\
	\S\ref{sect:mpl}				&	$10^{19}$ - $10^{24}$ g	&	3			&	2 au		&	$\infty$			&	1		&	0.01			&	Impact jetting	\\
	\S\ref{sect:f_d} 				&	$10^{23}$ g			&	1-10			&	2 au		&	$\infty$			&	1		&	0.01			&	Impact jetting	\\
	\S\ref{sect:r}				&	$10^{23}$ g			&	3			&	1 - 2.5 au	&	$\infty$			&	1		&	0.01			&	Impact jetting	\\	
	\S\ref{sect:tau_g} 			&	$10^{23}$ g			&	3			&	2 au		&	$10^6$ yr -$\infty$	&	1		&	0.01			&	Impact jetting	\\
	\S\ref{sect:f_acc} 			&	$10^{23}$ g			&	3			&	2 au		&	$\infty$			&	0.3 - 10	&	0.01			&	Impact jetting	\\
	\S\ref{sect:others} 			&	$10^{23}$ g			&	3			&	2 au		&	$\infty$			&	1		&	0.01 - 0.10	&	Impact jetting	\\
	\S\ref{sect:others} 			&	$10^{23}$ g			&	3			&	2 au		&	$\infty$			&	1		&	0.01			&	
	Impact jetting, constant production rate, decreasing production rate 
	\enddata 
\end{deluxetable*}

We perform simulations of chondrule formation and accretion, in which all the models are combined, following \S \ref{sect:model}.
In other words, the growth of a protoplanet, formation of chondrules, accretion of them both by the protoplanet and by planetesimals are computed simultaneously.
At first, we discuss the procedure of our simulations.
Then, chondrule formation and accretion in our fiducial model are presented. 
We explore the parameter dependences of $M_{\rm acc}$ and $\tau_{\rm acc}$. 
The parameter ranges in each model are summarized in Table \ref{table:models}.

\subsection{Synthesis}

Our simulations are composed of the growth of a protoplanet, chondrule formation, and chondrule accretion by the protoplanet and planetesimals. 
To synthesize these effects, we perform simulations based on the following procedure.
The mass of a protoplanet is increased by $dM$, which is calculated by equation (\ref{eq:dM_dt}) until its isolation mass, in a time interval $dt$.
After $v_{\rm imp}$ reaches 2.5 km s$^{-1}$, $F_{\rm ch}dM$ chondrules are formed in $dt$.
These chondrules are dealt as field chondrules.
The mass of field chondrules ($M_{\rm ch}$) is the sum of the remaining field chondrules in the previous step and $F_{\rm ch}dM$. 
Field chondrules are accreted by the protoplanet and planetesimals.
The chondrule mass accreted by the protoplanet in $dt$ (${\dot M_{\rm acc,pr}}dt$) is given by equation (\ref{eq:dM_acc_pr}). 
The chondrule mass by planetesimals (${\dot M_{\rm acc,pl}}dt$) depends on the accretion mode of planetesimals in each mass range (see \S \ref{sect:chondrule_acc_pl}).
The mass of the remaining field chondrules is given by $M_{\rm ch}-({\dot M_{\rm acc,pr}}+\sum {\dot M_{\rm acc,pl}} )dt$.
Then, a sequence of processes that can occur in a timestep ($dt$) are ended.
These processes are repeated until $3\times10^6$ yr to assess chondrule formation and accretion.

Note that while ${\dot M_{\rm acc,pr}}$ and ${\dot M_{\rm acc,pl}}$ are calculated independently, they are computed from the same amount of field chondrules. 
Some parameters affect either $\tau_{\rm acc,pr}$ or $\tau_{\rm acc,pl}$, but not both.
In such a case, both $M_{\rm acc,pr}$ and $M_{\rm acc,pl}$ are changed, since $M_{\rm ch}$ is changed.

\subsection{Fiducial model}\label{sect:fiducial}

\begin{figure}[ht]
	\plotone{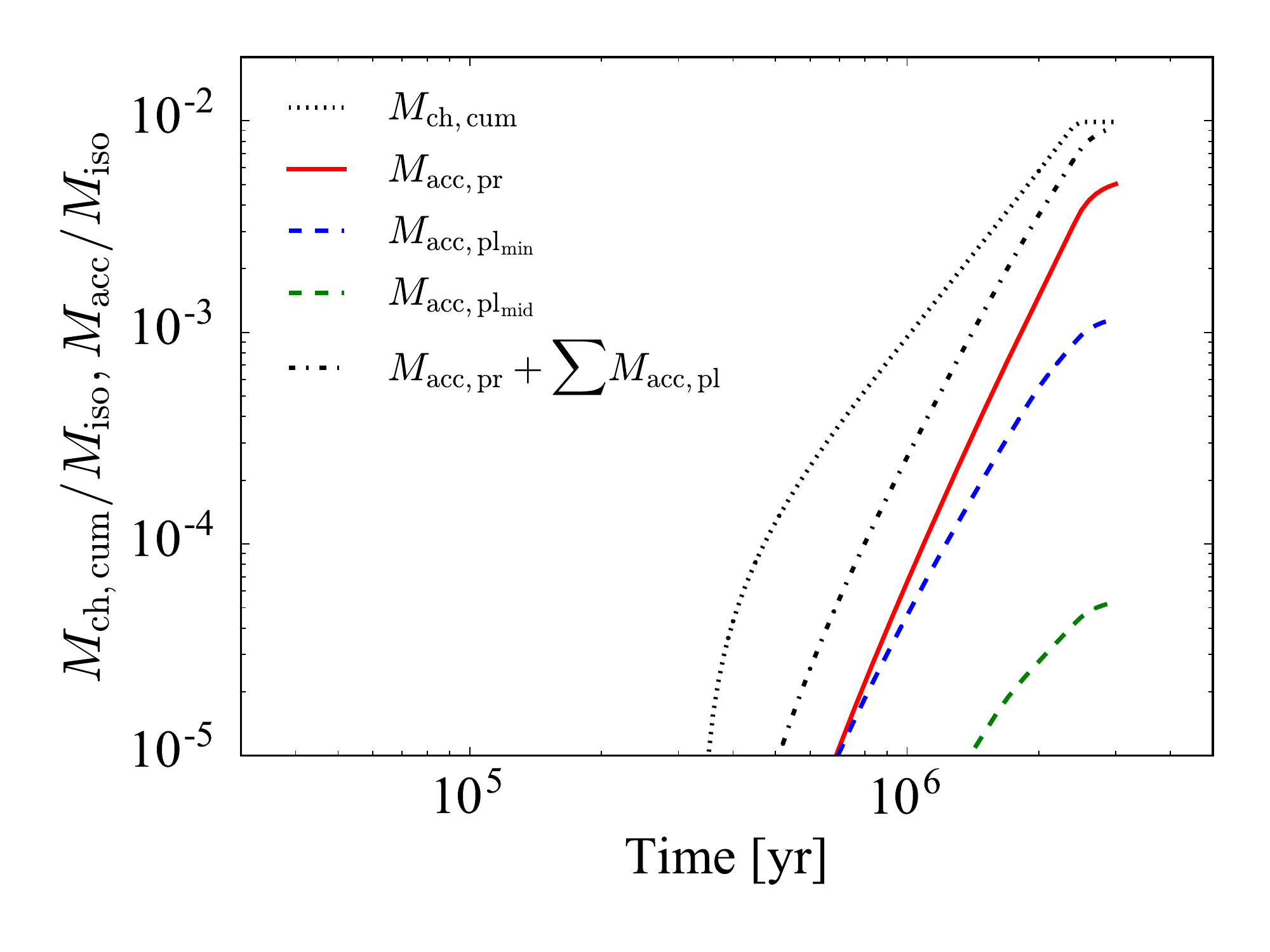}
	\caption{
		Time evolution of $M_{\rm ch,cum}$ (the dotted line, which is the same as the solid line in the bottle panel of Figure \ref{Fig:evolution_sample}), 
		 chondrules accreted by the protoplanet (solid line), planetesimals (dashed line), and all bodies (dot-dashed line). 
		In this figure, $M_{\rm acc,pl}$ of only pl$_{\rm min}$ planetesimals and pl$_{\rm mid}$ planetesimals are shown.
	}
	\label{Fig:t_Mch}
\end{figure}

Figure \ref{Fig:t_Mch} shows the mass of cumulative formed chondrules ($M_{\rm ch, cum}$)
 and those accreted by a protoplanet ($M_{\rm acc,pr}$) and planetesimals ($M_{\rm acc,pl}$) as a function of time.
The protoplanet accretes the largest amount of chondrules, and finally it accretes $5.0\times 10^{-3} M_{\rm iso}$, which is equal to 51\% of the formed chondrules (see the red, solid line). 
The smallest mass planetesimals have the second largest amount of the chondrules (see the blue, dashed line). 
They finally have $1.2\times 10^{-3}M_{\rm iso}$, which is 12\% of them. 
The chondrule mass accreted by all the planetesimals in single mass bins becomes smaller as $m_{\rm pl}$ increases, since $\tau_{\rm acc,pl}$ becomes longer (\S \ref{sect:chondrule_acc_pl}).
The summation of chondrules that are accreted by all the planetesimals in all the mass bins is 44\% of the formed chondrules. 
The most of the formed chondrules are accreted by the protoplanet and planetesimals (see the dot-dashed line).

We find that chondrules are not accreted soon after they formed. 
This is simply because the accretion timescale is $\gtrsim 10^5$ yr, which is much longer than the timescale of a collision, even for a protoplanet (see Figure \ref{Fig:M_vs_tau_acc}).
This feature can also be seen in Figure \ref{Fig:t_Mch}; for a given value of chondrule mass ($M_{\rm ch,cum}$ and $M_{\rm acc}$),
 there is a time-lag for the mass of chondrules accreted by all bodies (the dot-dashed line) to catch up with the cumulative value (the dotted line). 
This time-lag roughly corresponds to the accretion timescale of chondrules. 
Our results thus suggest that chondrules should have stayed in the solar nebula for 0.1 - 1 Myr. 
It is interesting that this time interval is roughly consistent with the isotope analysis of chondrules \citep{Akaki+2007}. 
In their study, the so-called compound chondrules, which are aggregates of two or more chondrules, were isotopically analyzed. 
They found that the secondary melting events occurred about 1 Myr after the primary melting happened. 
This infers that some of chondrules kept staying in the solar nebula for about 1 Myr.

\subsection{The dependence on $m_{\rm pl,min}$}\label{sect:mpl}

\begin{figure}[ht]
	\plotone{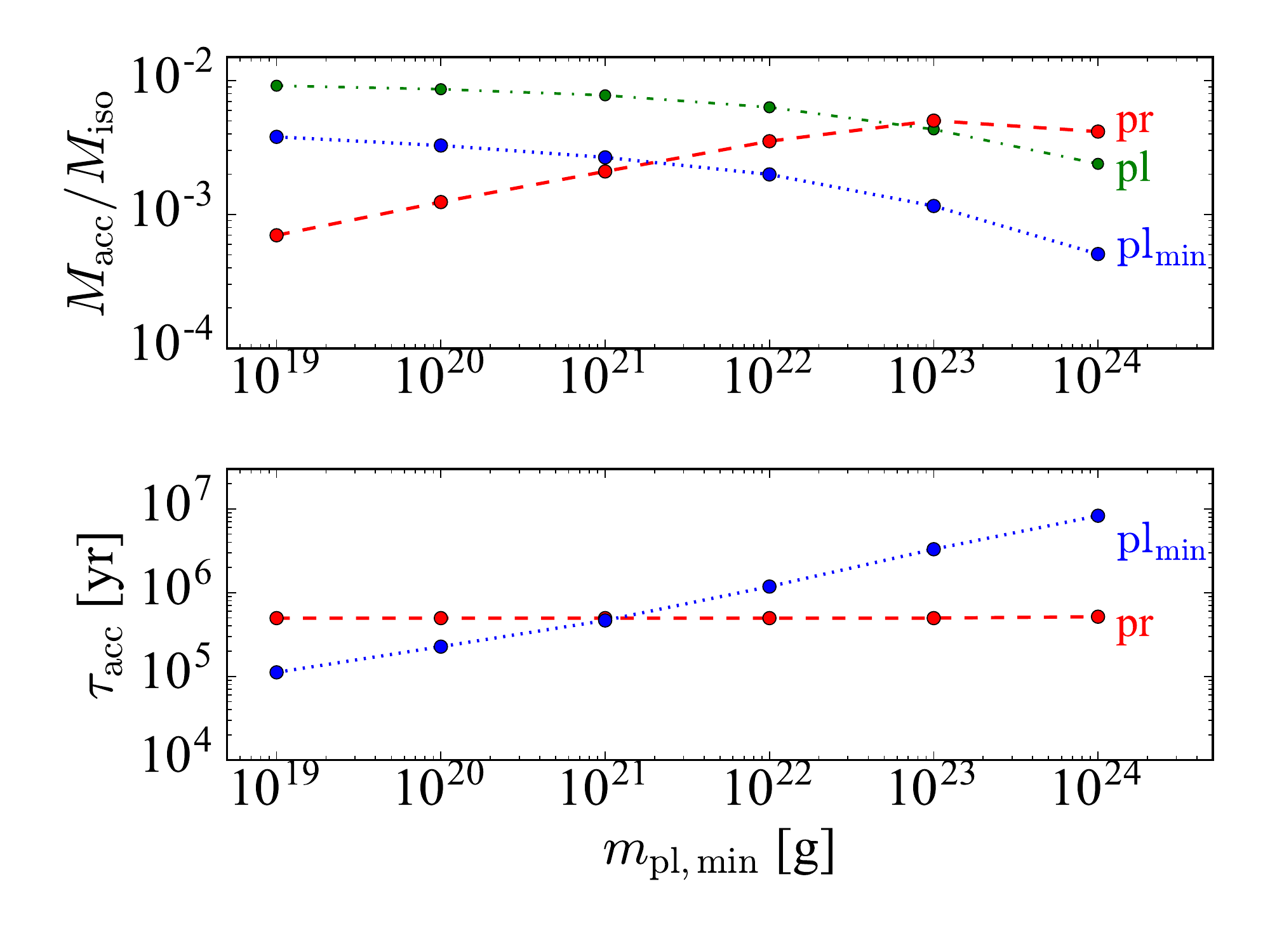}
	\caption{
		This figure shows $M_{\rm acc}$ (top) and $\tau_{\rm acc}$ (bottom) of a protoplanet (pr), the summation of all planetesimals (pl)
		  and pl$_{\rm min}$ planetesimals (pl$_{\rm min}$) at $3\times10^6$ yr as a function of $m_{\rm pl, min}$.
		The other parameters are listed as \S\ref{sect:mpl} in Table \ref{table:models}.
	}
	\label{Fig:mpl_vs_Macc+tau_acc}
\end{figure}

In this section, we examine the effect of planetesimal mass on chondrule formation and accretion.
We change the mass of the smallest planetesimals ($m_{\rm pl,min}$) and then perform similar simulations. 
Figure \ref{Fig:mpl_vs_Macc+tau_acc} shows $M_{\rm acc}$ and $\tau_{\rm acc}$ of a protoplanet and pl$_{\rm min}$ planetesimals
 at $3\times10^6$ yr as a function of $m_{\rm pl,min}$. 
The timescale of chondrule accretion by the protoplanet is constant with changing $m_{\rm pl,min}$ since it is independent of $m_{\rm pl, min}$ (equation \ref{eq:tau_acc_pr}). 
The timescale by pl$_{\rm min}$ planetesimals increases with increasing $m_{\rm pl,min}$. 
Considering $m_{\rm pl}=m_{\rm pl,min}$, equation (\ref{eq:tau_acc_pl_2}) is proportional to $m_{\rm pl,min}^{4/15}$.
This dependence comes from the product of $n_{\rm pl,min}\propto m_{\rm pl,min}^{-1}$, $e_{\rm pl,min}\propto m_{\rm pl,min}^{1/15}$,
 and $r_{\rm acc}^2\sim R_{\rm pl,min}^2\propto m_{\rm pl,min}^{2/3}$.
However, Figure \ref{Fig:mpl_vs_Macc+tau_acc} shows that $\tau_{\rm acc,pl}$ of pl$_{\rm min}$ planetesimals changes more rapidly than $m_{\rm pl,min}^{4/15}$.
This arises because the accretion timescale is additionally affected by the effect of the inclination ($f_{i_{\rm pl}}$) when $m_{\rm pl,min}>10^{21}$ g. 
In the case of $m_{\rm pl,min}\leq10^{21}$ g, $i_{\rm pl,min}$ is smaller than $h_{\rm ch}/r$, even when $M=M_{\rm iso}$,
 and $\tau_{\rm acc,pl}$ of pl$_{\rm min}$ planetesimals changes according to $m_{\rm pl,min}^{4/15}$.

Figure \ref{Fig:mpl_vs_Macc+tau_acc} also shows that $M_{\rm acc,pr}$ increases as $m_{\rm pl,min}$ increases when $m_{\rm pl,min}<10^{24}$ g.
This occurs because $M_{\rm acc,pl}$ decreases as $m_{\rm pl,min}$ increases.
When ${\dot M_{\rm acc,pl}}$ becomes smaller, more chondrules remain as field chondrules in a step.
Since the mass of field chondrules increases, the chondrule accretion rate by a protoplanet (${\dot M_{\rm acc,pr}}=M_{\rm ch}/\tau_{\rm acc,pr}$) becomes larger at the subsequent timesteps.
On the contrary, $M_{\rm acc,pr}$ at $m_{\rm pl,min}=10^{24}$ g becomes smaller than that at $m_{\rm pl,min}=10^{23}$ g.
When $m_{\rm pl,min}\geq10^{24}$ g, the mass of the protoplanet does not reach $M_{\rm iso}$ within $3 \times 10^6$ yr,
 since $\tau_{\rm pr}$ becomes larger due to larger $e_{\rm pl,min}$ (see equation \ref{eq:tau_pr}).
Then, the cumulative formed chondrules mass is smaller than $F_{\rm ch}M_{\rm iso}$. 
Since the total mass of chondrules decreases, $M_{\rm acc,pr}$ also decreases.
As increasing $m_{\rm pl,min}$, $M_{\rm acc,pl_{\rm min}}$ decreases due to the increase of $\tau_{\rm acc,pl}$ (see equation \ref{eq:tau_acc_pl_2}, also see Figure \ref{Fig:M_vs_tau_acc}). 
Except for $m_{\rm pl,min}=10^{24}$ g, the protoplanet accretes $0.018M_{\rm iso}$ - $0.050M_{\rm iso}$, which is equal to 19\% - 50\% of the formed chondrules. 
On the other hand, planetesimals accrete 44\% - 81\% of the formed chondrules in total. 
The smallest planetesimals get the larger amount of chondrules in planetesimals, which is 12\% - 28\% of the formed chondrules. 

\subsection{The dependence on $f_{\rm d}$}\label{sect:f_d}

\begin{figure}[ht]
	\plotone{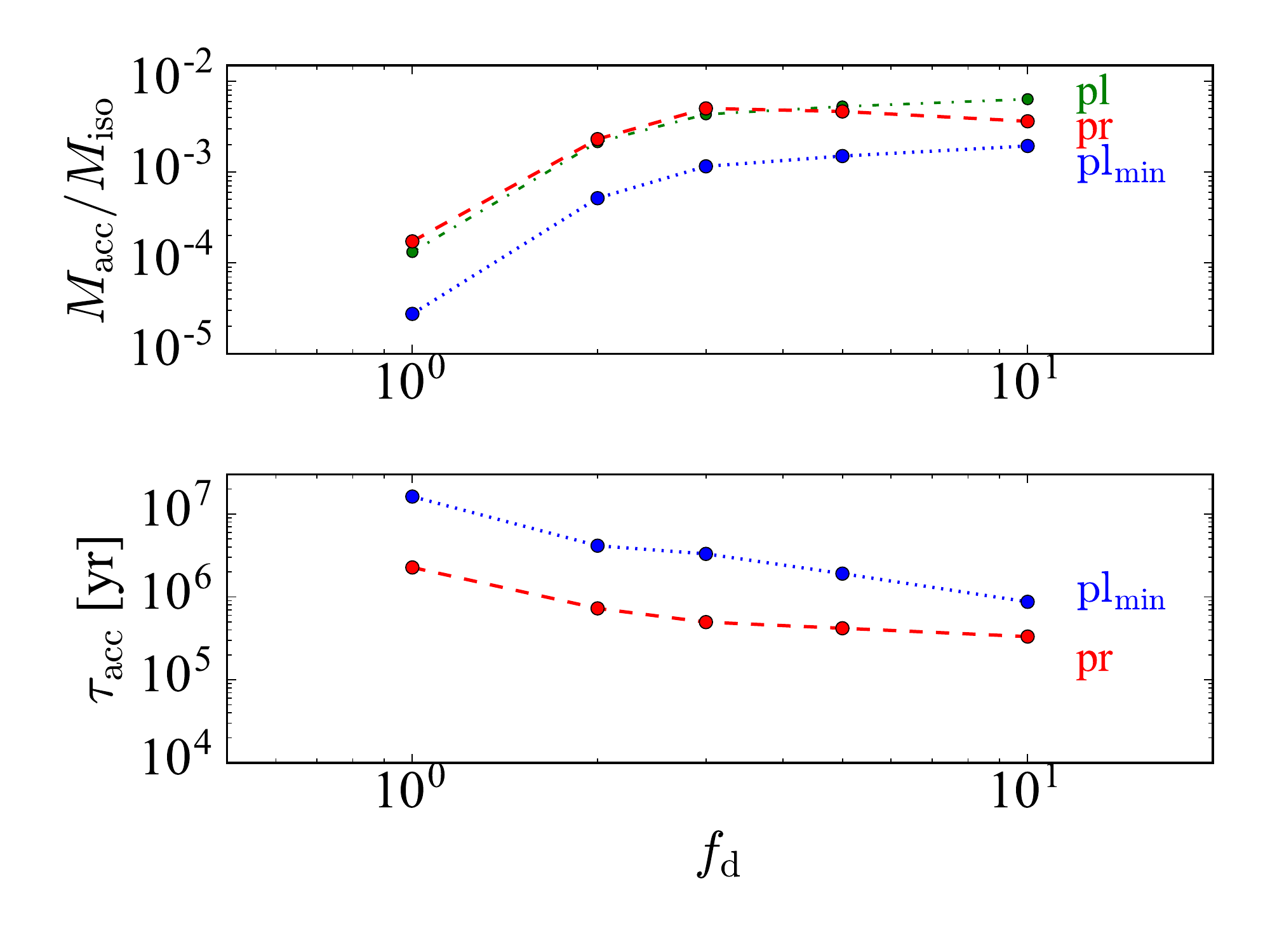}
	\caption{ 
		Same as Figure \ref{Fig:mpl_vs_Macc+tau_acc}, but for a function of $f_{\rm d}$.		
		Note that the mass range (the top panel) is expanded to $10^{-5}$ - $10^{-2}$, compared with other figures.
		The mass of the pl$_{\rm min}$ planetesimal is $10^{23}$ g, and the other parameters are listed as \S\ref{sect:f_d} in Table \ref{table:models}.
		}
	\label{Fig:fd_vs_Macc+tau_acc}
\end{figure}

\cite{Hasegawa+2016b} showed that there are appropriate values of $f_d$ and $m_{\rm pl}$ for chondrule formation and accretion by the impact jetting process.
In this section, we examine how the timescale of chondrule accretion and amount of accreted chondrules depend on $f_{\rm d}$.
We adopt $f_{\rm d}$ = 1, 2,3 (fiducial), 5, and 10. 
Figure \ref{Fig:fd_vs_Macc+tau_acc} shows $M_{\rm acc}$ and $\tau_{\rm acc}$ as a function of $f_{\rm d}$. 
Note that $M_{\rm iso}$ is proportional to $f_{\rm d}^{3/2}$.
As $f_{\rm d}$ increases, $\tau_{\rm acc,pr}$ and $\tau_{\rm acc,pl}$ become shorter. 
The protoplanet does not reach its isolation mass within $3\times10^6$ yr, when $f_{\rm d}<2.7$ (equation \ref{eq:tau_pr}).
This is why $\tau_{\rm acc}$ of the protoplanet and pl$_{\rm min}$ planetesimals inflects around $f_{\rm d}=3$.
Since $\tau_{\rm pr}\propto m_{\rm pl,min}^{2/15} f_{\rm d}^{-9/10}$, the protoplanet can get $M_{\rm iso}$ if $f_{\rm d}=1$ and
 $m_{\rm pl,min}\leq 1.2\times10^{20}$ g.

When the protoplanet gets $M_{\rm iso}$, the $f_{\rm d}$ dependence on $\tau_{\rm acc,pr}$ is caused
 by $r_{\rm acc}^2\propto f_{\rm d}M_{\rm iso}^{-1}\propto f_{\rm d}^{-1/2}$.
The dependence on $\tau_{\rm acc,pl_{\rm min}}$ is
 $\tau_{\rm acc,pl_{min}}\propto n_{\rm pl} r_{\rm acc}^2 e_{\rm pl} f_{i_{\rm pl}}^{-1}\propto f_{\rm d}^{-13/10}$.
The dependence of $\tau_{\rm acc,pl_{min}}$ on $f_{\rm d}$ is stronger than that of $\tau_{\rm acc,pr}$.
Then $M_{\rm acc,pl_{min}}/M_{\rm iso}$ becomes larger as $f_{\rm d}$ increases.

It is important that the mass ratio of accreted chondrules between a protoplanet and planetesimals does not change very much when $f_{\rm d} > 3$.
Even for the case of $f_{\rm d} < 3$, the trend of our results does not change; most chondrules are accreted by a protoplanet.  
Thus, the results obtained from our fiducial case can be applicable for a wide range of disk masses.

\subsection{The dependence on $r$}\label{sect:r}

\begin{figure}[ht]
	\plotone{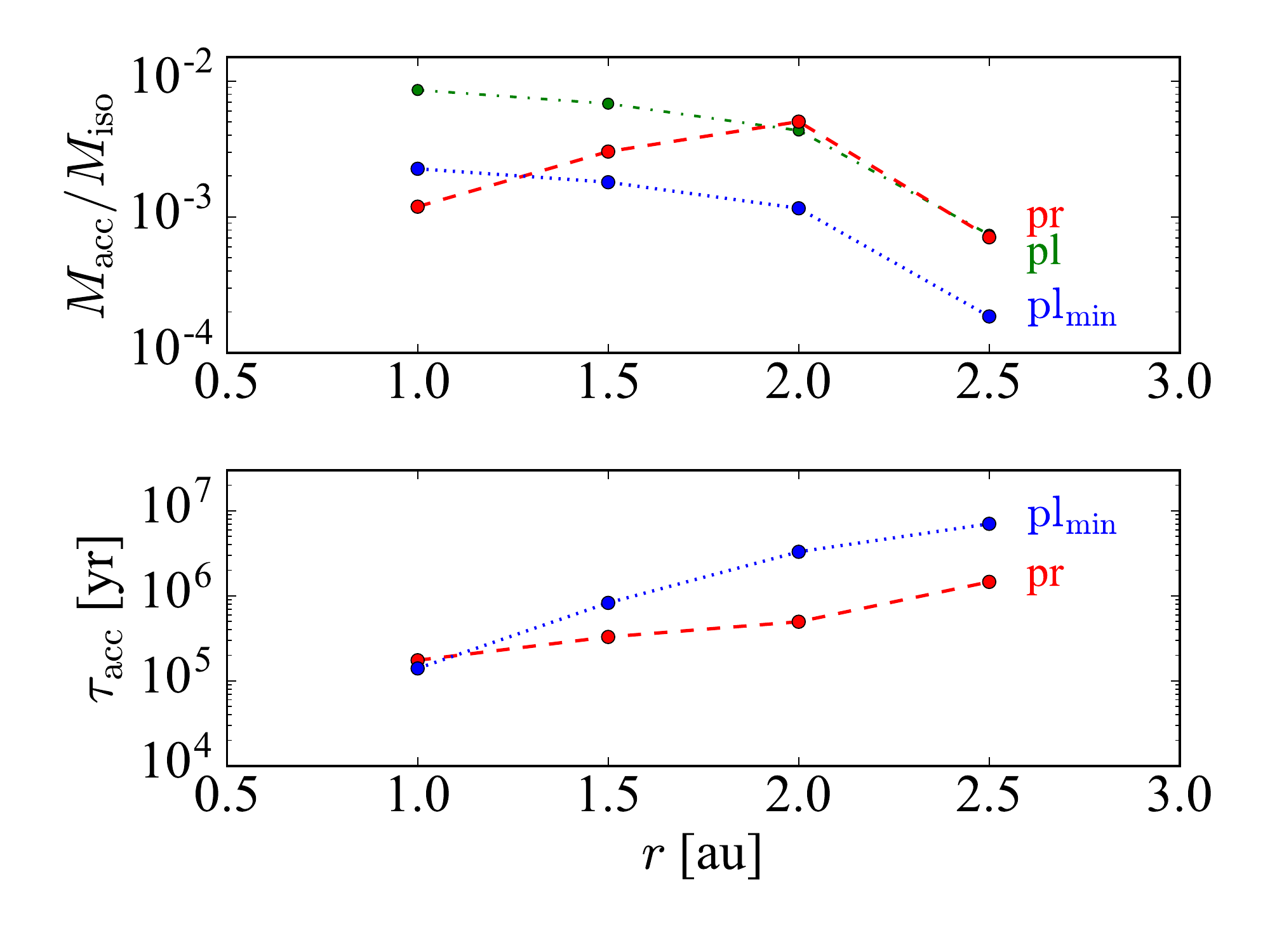}
	\caption{
		Same as Figure \ref{Fig:mpl_vs_Macc+tau_acc}, but for a function of $r$.
		The mass of the pl$_{\rm min}$ planetesimal is $10^{23}$ g, and the other parameters are listed as \S\ref{sect:r} in Table \ref{table:models}.
		}
	\label{Fig:r_vs_Macc+tau_acc}
\end{figure}

The orbital radius varies the timescales of chondrule formation and accretion.
We perform simulations with changing orbital radii from 1.0 au to 2.5 au. 
The timescale of chondrule accretion becomes longer as $r$ increases. 
Figure \ref{Fig:r_vs_Macc+tau_acc} shows $M_{\rm acc}$ and $\tau_{\rm acc}$ of a protoplanet and pl$_{\rm min}$ planetesimals
 at $3\times10^6$ yr as a function of $r$. 
The accreted chondrules by both the protoplanet and by pl$_{\rm min}$ planetesimals drop at $r=2.5$ au (see the top panel).
This is because the protoplanet does not reach $M_{\rm iso}$ within in $3 \times 10^6$ yr, as discussed in the above section. 
In the following, we consider chondrule accretion at $r<2.5$ au. 

Based on the derivation in Section \ref{sect:model}, $\tau_{\rm acc,pr} \propto r^{3/2}$ under the approximation of $H^2/(1+H^2)\sim H^2$ (equation \ref{eq:tau_acc_pr}),
 while $\tau_{\rm acc,pl_{\rm min}}$ changes more rapidly, which is given as $\tau_{\rm acc,pl_{min}}\propto r^{24/5}$ (equation \ref{eq:tau_acc_pl_2}). 
This indicates that as $r$ decreases, both $\tau_{\rm acc}$ of the protoplanet and the pl$_{\rm min}$ planetesimals decrease. 
We find that $\tau_{\rm acc,pl_{\rm min}}\simeq \tau_{\rm acc,pr}$ when $3\times10^6$ yr at 1.0 au (see Figure \ref{Fig:r_vs_Macc+tau_acc}). 
Then, more chondrules are accreted by pl$_{\rm min}$ planetesimals than by the protoplanet at 1.0 au. 
For this case, the protoplanet accretes about 12\% of the formed chondrules and 88\% of them are accreted by planetesimals. 
Although $\tau_{\rm acc,pl_{\rm min}}$ at 1.0 au is about 20 times shorter than that at 2.0 au,
 $M_{\rm acc,pl_{\rm min}}$ at 1.0 au is 23 \% of formed chondrules, which is only twice larger than that at 2.0 au.
In other words, $M_{\rm acc,pl}$ is relatively insensitive to the change of $\tau_{\rm acc,pl}$.
When $\tau_{\rm acc,pl}$ becomes small, chondrules are more quickly accreted, and $M_{\rm ch}$ becomes smaller at the same time.
Then, the final values of $M_{\rm acc}$, which are given by $\int {\dot M_{\rm acc}} dt=\int (M_{\rm ch}/\tau_{\rm acc})dt$, are not proportional to $\tau_{\rm acc}^{-1}$.

\subsection{The dependence on $\tau_{\rm g}$}\label{sect:tau_g}

The above simulations are performed without gas depletion. 
When the gas density and surface density are changed with time, $\tau_{\rm stop}$ and $e_{\rm pl}$ also vary. 
Since we give gas depletion by $\exp{(-t/\tau_{\rm g})}$, $\tau_{\rm stop}$ and $e_{\rm pl}$ increase as gas disks evolve with time;
 $\tau_{\rm stop}\propto \rho_{\rm g}^{-1} \propto \exp{(t/\tau_{\rm g})}$ (equation \ref{eq:tau_stop}),
 and $e_{\rm pl} \propto \rho_{\rm g}^{-1/5} \propto \exp{(0.2t/\tau_{\rm g})}$ (equation \ref{eq:e_pl}).
This means that when $\tau_{\rm g}\gtrsim t_{\rm iso}$, which is $2.4\times10^6$ yr in our fiducial model (\S \ref{sect:chondrule_formation}),
 $\tau_{\rm stop}$ and $e_{\rm pl}$ are changed only by a factor of a few.
Note that $H$ does not depend on $\tau_{\rm g}$, since the $\tau_{\rm g}$ dependence is cancelled,
 due to $H\propto (\alpha_{\rm eff} / \tau_{\rm stop} )^{1/2} \propto (\Sigma_{\rm g}^{-1}/\rho_{\rm g}^{-1})^{1/2}$ (equations \ref{eq:alpha_eff} and \ref{eq:H}).

\begin{figure}[ht]
	\plotone{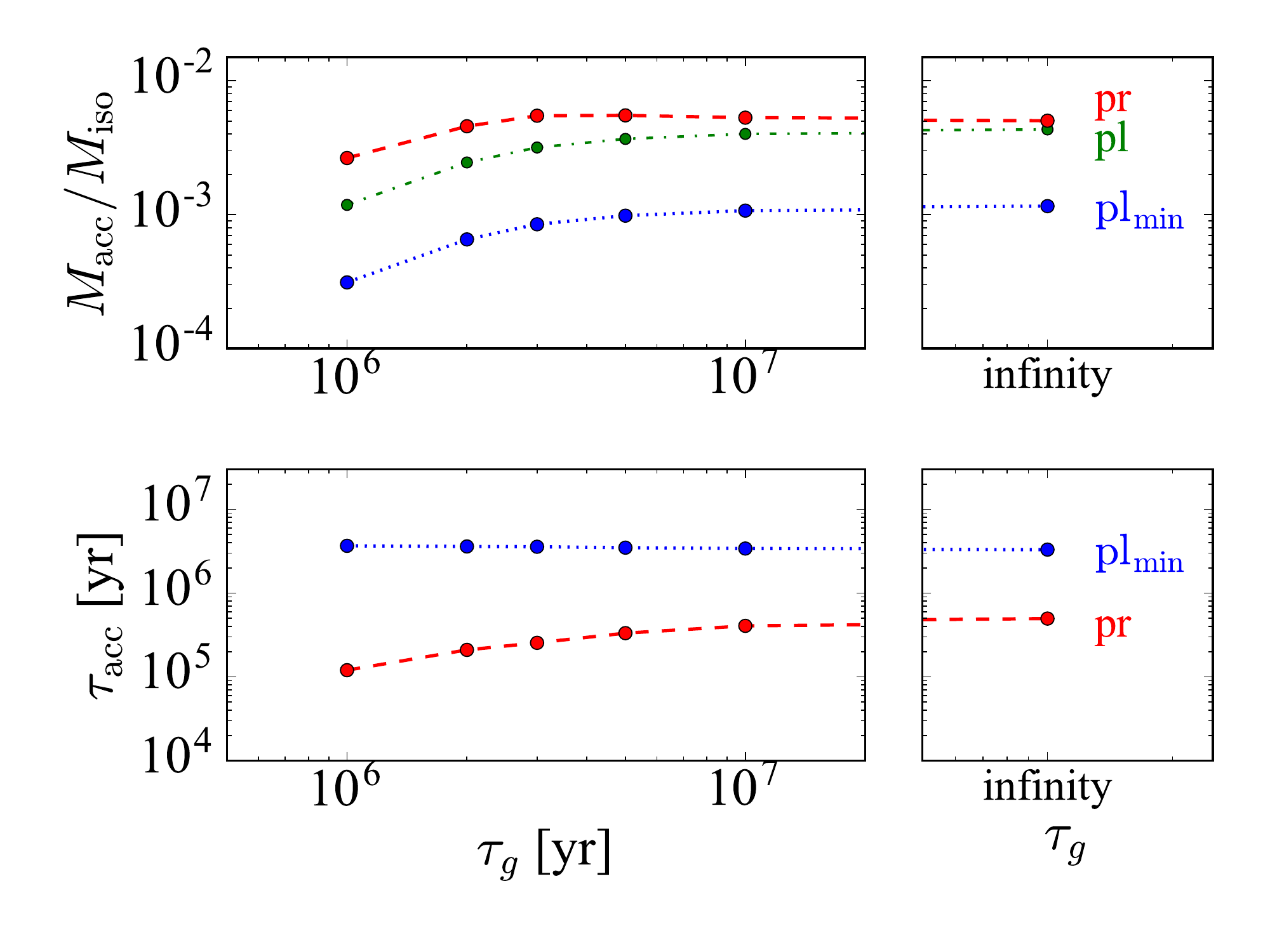}
	\caption{
		Same as Figure \ref{Fig:mpl_vs_Macc+tau_acc}, but for a function of $\tau_{\rm g}$.
		Note that these are the values at $3\times10^6$ yr. 
		The mass of the pl$_{\rm min}$ planetesimal is $10^{23}$ g, and the other parameters are listed as \S\ref{sect:tau_g} in Table \ref{table:models}.
	}
	\label{Fig:taug_vs_Macc+tau_acc}
\end{figure}

We perform simulations with $\tau_{\rm g}=10^6$ yr, $2\times10^6$ yr, $3\times10^6$ yr, $5\times10^6$ yr, and $10^7$ yr.
Note that while we consider the cases of $\tau_{\rm g} = 10^6$ yr and $2 \times 10^6$ yr only for completeness,
 the results for the case of $\tau \ge 3$ Myr are more appropriate for chondrules found in chondrites. 
This is because chondrule formation likely continued until 3 Myr after CAI formation,
 and a gas disk would be needed for chondrule formation at that time \citep[e.g.,][]{Hewins+2005}.
Our fiducial model can be viewed as $\tau_{\rm g}=$infinity.
Figure \ref{Fig:taug_vs_Macc+tau_acc} shows the resultant values of $M_{\rm acc}$ and $\tau_{\rm acc}$ for the protoplanet and pl$_{\rm min}$ planetesimals at $3\times10^6$ yr.
As $\tau_{\rm g}$ increases, $\tau_{\rm acc,pr}$ increases, and $\tau_{\rm acc,pl}$ is hardly changed.
The $\tau_{\rm g}$ dependence on $\tau_{\rm acc}$ arises from $r_{\rm acc}^{-2} \Delta v^{-1}$.
In the case of the protoplanet, $r_{\rm acc}^{-2} \Delta v^{-1} \propto \tau_{\rm stop}^{-1} \propto \exp{(- t/\tau_{\rm g})}$ (equations \ref{eq:tau_stop} and \ref{eq:r_acc_pr}).
This is why $\tau_{\rm acc,pr}$ increases with increasing $\tau_{\rm g}$ under $\tau_{\rm g}\gtrsim t_{\rm iso}$.
For pl$_{\rm min}$ planetesimals, $\tau_{\rm acc,pl}$ is multiplied by $f_{i_{\rm pl}}^{-1}$ at $t=3\times 10^6$ yr. 
Since $\tau_{\rm acc,pl}\propto r_{\rm acc}^{-2} \Delta v^{-1}f_{i_{\rm pl}}^{-1}$, which is approximately proportional to $ i_{\rm pl}/e_{\rm pl}$,
 $\tau_{\rm acc,pl}$ does not depend on $\tau_{\rm g}$.

For this case, $M_{\rm acc}$ of the protoplanet and pl$_{\rm min}$ planetesimals keep similar values, compared with the fiducial case.
When $\tau_{\rm g}\lesssim t_{\rm iso}$, $M_{\rm acc}$ of them increases as $\tau_{\rm g}$ increases.
This is because gas depletion occurs before the protoplanet reaches its isolation mass. 
Due to the gas depletion, $e_{\rm pl}$ increases, and $\tau_{\rm pr}$ becomes longer (equations \ref{eq:dM_dt} and \ref{eq:tau_pr}).
The protoplanet does not get $M_{\rm iso}$ in $3\times10^6$ yr, and $M_{\rm acc}$ becomes a small value, since cumulative formed chondrules mass becomes small.

\subsection{The dependence on $f_{\rm acc}$}\label{sect:f_acc}

\begin{figure}[ht]
	\plotone{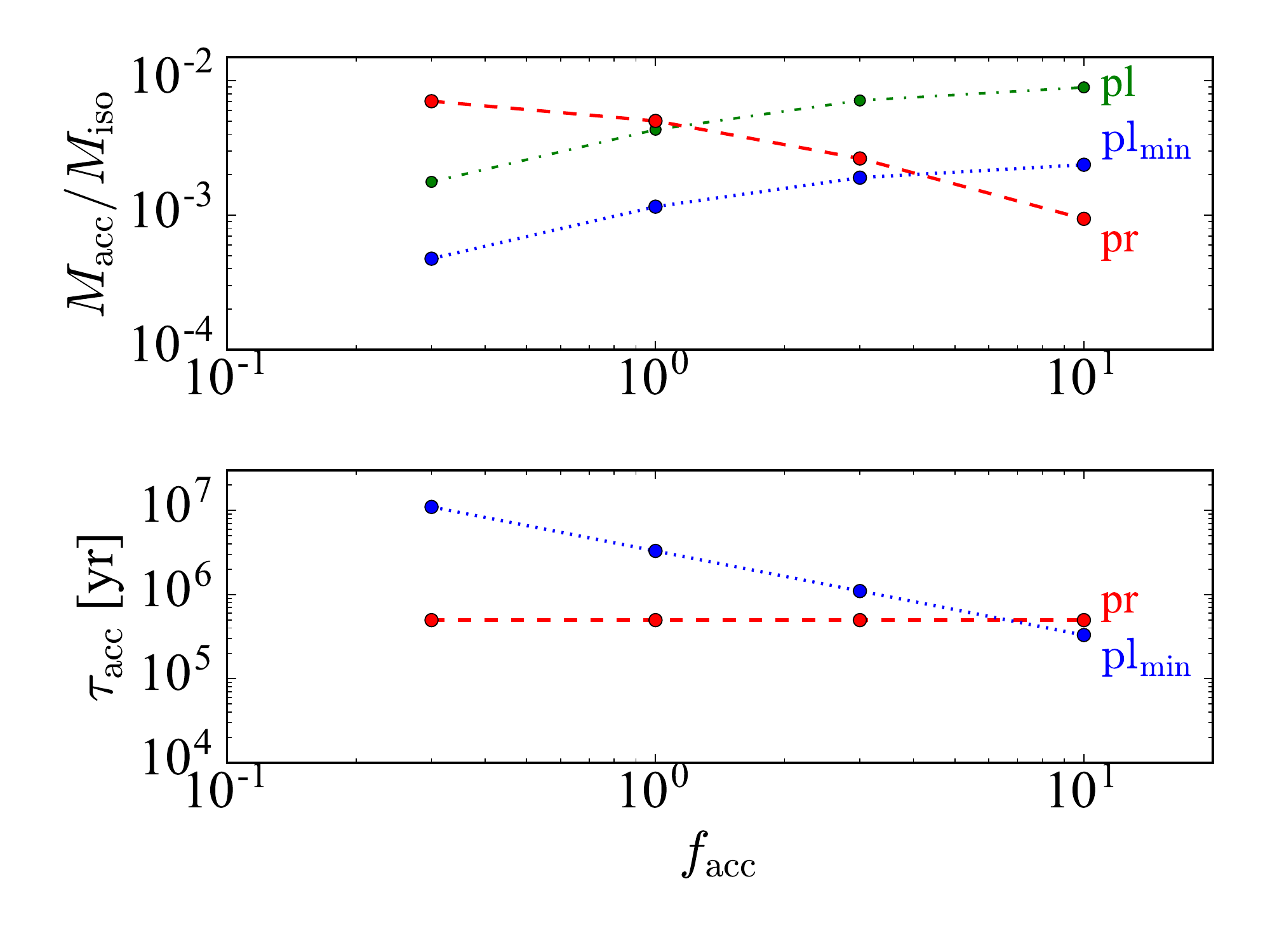}
	\caption{
		Same as Figure \ref{Fig:mpl_vs_Macc+tau_acc}, but for a function of $f_{\rm acc}$.
		The mass of the pl$_{\rm min}$ planetesimal is $10^{23}$ g, and the other parameters are listed as \S\ref{sect:f_d} in Table \ref{table:models}.
		}
	\label{Fig:facc_vs_Macc+tau_acc}
\end{figure}

In this paper, our model is developed, based on the oligarchic growth model in laminar disks \citep{Kokubo&Ida2000}.
As described in \S \ref{sect:timescale_chondrule_acc}, chondrule accretion can be affected by disk turbulence. 
In this section, we multiply $\tau_{\rm acc,pl}$ by $f_{\rm acc}$ to consider the case of more effective accretion of chondrules, which can be triggered by disk turbulence.
We adopt $f_{\rm acc}=0.3$, 1 (fiducial), 3, and 10.
In these simulations, $\tau_{\rm acc,pl}\propto f_{\rm acc}^{-1}$, and $\tau_{\rm acc,pr}$ is constant with changing $f_{\rm acc}$ (see Figure \ref{Fig:facc_vs_Macc+tau_acc}).
Our results show that the chondrule mass accreted by pl$_{\rm min}$ planetesimals does not change in proportional to $f_{\rm acc}$ (see Figure \ref{Fig:facc_vs_Macc+tau_acc}). 
As we see in \S \ref{sect:r}, the dependence of $M_{\rm acc,pl}$ on $\tau_{\rm acc,pl}$ is weak, since ${\dot M_{\rm acc}}=M_{\rm ch}/\tau_{\rm acc}$,
 and $M_{\rm ch}$ becomes smaller when $\tau_{\rm acc,pl}$ becomes small.
As a result, the $M_{\rm acc,pl_{\rm min}}$ dependence on $f_{\rm acc}$ becomes small,
 and pl$_{\rm min}$ planetesimals accrete 24\% of the formed chondrules even when $f_{\rm acc}=10$.

\subsection{The other dependences}\label{sect:others}

We also perform simulations with changing $F_{\rm ch}$ and chondrule formation models.
When we change $F_{\rm ch}$, the mass of the formed chondrule is changed in proportional to $F_{\rm ch}$. 
Since $\tau_{\rm acc}$ does not depend on $F_{\rm ch}$, $M_{\rm acc}$ of a protoplanet and planetesimals is proportional to $F_{\rm ch}$.

When we change chondrule formation models, we fix the timescale of chondrule formation (i.e., $M_{\rm esc}\leq M\leq M_{\rm iso}$)
 and the total mass of the formed chondrules (see Section \ref{sect:chondrule_formation}).
We perform simulations with the constant production rate model and decreasing production rate model (\S \ref{sect:chondrule_formation}).
The chondrule mass accreted by pl$_{\rm min}$ planetesimals increases in the following order, the impact jetting model (fiducial),
 the constant production rate model, and the the decreasing production rate model. 
This is because pl$_{\rm min}$ planetesimals accrete more chondrules than the protoplanet when $M\simeq M_{\rm esc}$ (see Figure \ref{Fig:M_vs_tau_acc}). 
However, the final chondrule mass accreted by pl$_{\rm min}$ planetesimals changes slightly, $1.2\times10^{-3}M_{\rm iso}$ in the impact jetting model,
 $1.3\times10^{-3}M_{\rm iso}$ in the constant production rate model, and $1.5\times10^{-3}M_{\rm iso}$ in the the decreasing production rate model. 
This arises because the condition that $\tau_{\rm acc,pl_{\rm min}}<\tau_{\rm acc,pr}$ is satisfied only for the initial $2\times10^5$ yr
 in the total chondrule forming timescale of $2 \times 10^6$ yr. 
And, $\tau_{\rm acc,pl_{\rm min}}$ in this initial time interval is about $5\times10^6$ yr, which is quite long, compared with the interval.
This is why the resultant chondrule masses accreted by planetesimals become similar values for all the models. 
Thus, the chondrule formation models have little influence on our results.

\section{DISCUSSION}\label{sect:discussion}

\subsection{Chondrules on planetesimals}

A protoplanet accretes most chondrules in many simulations. 
It accretes about 50\% of the formed chondrules under the condition that $m_{\rm pl,min}\sim 10^{23}$ g, $f_{\rm d}=3$, $r\simeq 2$ au, and $f_{\rm acc}\simeq 1$. 
The remnant chondrules are accreted by planetesimals. 
For planetesimals, the smallest planetesimals accrete the largest mass of them. 
Under the above condition, pl$_{\rm min}$ planetesimals finally get about 10\% of the formed chondrules.
The other 40\% of them are accreted by the other planetesimals.

These chondrules would not contribute to planetesimal growth. 
The planetesimals in each mass bin get $f_{{\rm r},i} F_{\rm ch} M_{\rm iso}$ ($f_{{\rm r},i}\leq f_{\rm r,min}\simeq0.1$) chondrules, 
 where $f_{{\rm r},i}$ is the mass fraction of chondrule accreted by planetesimals in $i$-th mass bin
 and given by $f_{{\rm r},i}= M_{{\rm acc,pl}_i} / ( M_{\rm acc,pr} + \sum M_{\rm acc,pl} )\simeq M_{{\rm acc,pl}_i}/F_{\rm ch} M_{\rm iso}$.
The mass fraction of chondrules with respect to an accreting planetesimal is 
\begin{eqnarray}
	f_{\rm m,ch} &=& \frac{f_{{\rm r},i} F_{\rm ch}M_{\rm iso}}{m_{\rm pl}n_{\rm pl}}  \nonumber \\
		&=& 6.0 \times10^{-2} f_{{\rm r},i} 
			\left( \frac{m_{\rm pl} }{10^{23}\mbox{ g}} \right)
			 \left( \frac{\Sigma_{\rm d}}{11\mbox{ g cm}^{-2}} \right)^{3/10} \nonumber \\ && \times 
			\left( \frac{r}{\mbox{2 au}} \right)^{3/5}  
			\left( \frac{m_{\rm pl,min}}{10^{23}\mbox{ g} } \right)^{-6/5},
\end{eqnarray}
where $\Sigma_{\rm d}$ is about 11 $\mbox{g cm}^{-2}$ at 2 au when $f_{\rm d}=3$. 
In the case of the smallest planetesimals, i.e., $m_{\rm pl}=m_{\rm pl,min}$, $f_{\rm r,min}\simeq 0.1$, we find that $f_{\rm m,ch}=6.0\times10^{-3}$,
 which means that mass of the accreted chondrules are much smaller than the planetesimal mass.
This equation is seemingly proportional to $m_{\rm pl}$. 
However, since $f_{{\rm r},i}$ decreases with increasing $m_{\rm pl}$ (see \S \ref{sect:fiducial} and Figure \ref{Fig:t_Mch}), $f_{\rm m,ch}$ keeps small values.
The dependence of $f_{{\rm r},i}$ on $m_{\rm pl}$ can be derived from $\tau_{\rm acc,pl}$.
Considering that planetesimals are in hyperbolic regime, $f_{{\rm r},i}\propto m_{\rm pl}^{-19/15}$ with the condition that $f_{i_{\rm pl}}=1$
 (see equation (\ref{eq:tau_acc_f_i})), we obtain that $f_{\rm m,ch}\propto m_{\rm pl}^{-4/15}$.
The small $f_{\rm m,ch}$ means that accreted chondrule does not change mass of planetesimals.

\begin{figure}[ht]
	\plotone{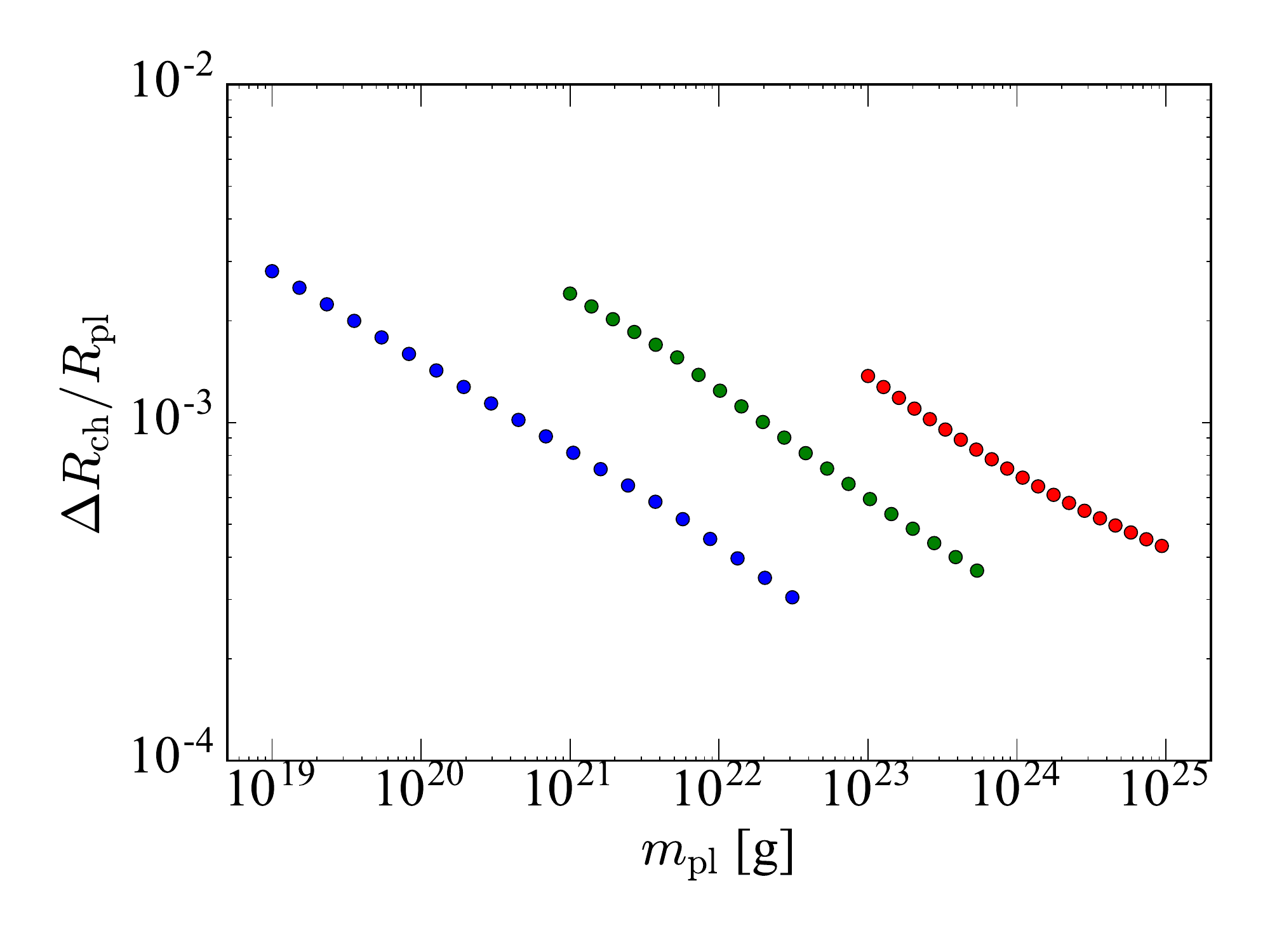}
	\caption{
	The thickness of chondrule layers on planetesimals normalized $R_{\rm pl}$ is shown as a function of $m_{\rm pl}$.
	The red dots represent $\Delta R_{\rm ch}/R_{\rm pl}$ in our fiducial model ($m_{\rm pl, min}=10^{23}$ g). 
	We also plot the results of models changing $m_{\rm pl,min}$, $m_{\rm pl,min}=10^{21}$ g (the green ones) and $10^{19}$ g (the blue ones).
	}
	\label{Fig:chondrule_layer}
\end{figure}

On the other hand, 
this fraction is too small to reproduce the fractional abundance of chondrules in chondrites \citep[e.g.,][]{Scott&Krot2005}.
In other words, when the current samples of chondrites originated from fragments of massive bodies,
 our results suggest that fragments arising only from planetesimals' surfaces can satisfy the measured abundance of chondrules in chondrites.
The accreted chondrules by planetesimals make a chondrule-rich layer in the surface region of planetesimals.
The thickness of this layer normalized by $R_{\rm pl}$ is computed as 
\begin{eqnarray}
	\frac{\Delta R_{\rm ch}}{R_{\rm pl}} &=& \frac{ f_{\rm m,ch}m_{\rm pl} }{ 4\pi R_{\rm pl}^3 \rho_{\rm s} } \nonumber \\
	&=& 1.2 \times 10^{-2} f_{{\rm r},i}  
		\left( \frac{\rho_{\rm pl}}{2 \mbox{ g cm}^{-3}} \right) 
		\left( \frac{\rho_{\rm s}}{3.3 \mbox{ g cm}^{-3}} \right)^{-1} \
		\nonumber \\ && \times 
		\left( \frac{m_{\rm pl} }{10^{23}\mbox{ g}} \right)
		 \left( \frac{\Sigma_{\rm d}}{11\mbox{ g cm}^{-2}} \right)^{3/10} 
		\left( \frac{r}{\mbox{2 au}} \right)^{3/5}   \nonumber \\ && \times 
		\left( \frac{m_{\rm pl,min}}{10^{23}\mbox{ g} } \right)^{-6/5} .
	\label{eq:R_ch}
\end{eqnarray}
Figure \ref{Fig:chondrule_layer} shows the results of $\Delta R_{\rm ch}/R_{\rm pl}$ as a function of $m_{\rm pl}$,
 which are obtained from our calculations of the accreted chondrule mass (see Section \ref{sect:mpl}). 
We find that for the case of $m_{\rm pl,min}=10^{21}$ g (see the green dots), the results are characterized well by $m_{\rm pl}^{-4/15}$
 while for the case of $m_{pl,min}=10^{19}$ g (see the blue dots), they are by $m_{\rm pl}^{-1/3}$. 
These can be explained by the behavior of $\Delta R_{\rm ch}/R_{\rm pl}$ ($\propto f_{{\rm r},i} m_{\rm pl}$);
 for the former case, $\Delta R_{\rm ch}/R_{\rm pl} \propto m_{\rm pl}^{-4/15}$ under the condition that $f_{i_{\rm pl}}=1$. 
For the latter one, $\Delta R_{\rm ch}/R_{\rm pl}\propto m_{\rm pl}^{-1/3}$ when $f_{i_{\rm pl}}=1$ is given by equation (\ref{eq:tau_acc_f_i}). 
Our results also show that for the case of $m_{\rm pl,min}=10^{23}$ g, the dependence of $\Delta R_{\rm ch}/R_{\rm pl}$ on $m_{\rm pl}$
 is weaker than $m_{\rm pl}^{-4/15}$, since larger mass planetesimals are in settling regime.
In the case of $m_{\rm pl,min}=m_{\rm pl}=10^{23}$ g, which are planetesimals with the radius of 230 km, the planetesimal has the 0.27 km chondrule layer on its surface.
\cite{Wakita+2016b} showed that the majority of ejecta arise from a very thin surface layer, which is about 100 m from the surface.
Then, the (high) abundance of chondrules in chondrites can be potentially explained by the chondrule layer if the original materials of chondrites are in this layer. 
Based on a high fractional abundance of chondrules in chondrites, it can be expected
 that there was not a large amount of dust, which has a similar Stokes number to chondrules in the solar nebula at that time.

\subsection{Other effects}

In our simulations, we assume that chondrules stay at their formed orbits. 
Theoretical studies suggested that chondrules migrate inward due to gas drag.
This migration timescale is $\sim 10^5$ yr \citep{Adachi+1976,Weidenschilling1977}.
This timescale is shorter than the timescales of chondrule accretion (Figure \ref{Fig:t_Mch}),
 which indicates that chondrule would migrate inward before they are accreted by a protoplanet and planetesimals.
On the other hand, the isotopic measurement of compound chondrules suggested that the chondrules stayed in the solar nebula for 1 Myr \citep{Akaki+2007}.
Some mechanism, such as a radial pressure bump \citep[e.g.,][]{Taki+2016} or vortices \citep[e.g.,][]{Cuzzi+2010,Fu_W+2014},
 would be needed to have kept chondrules from migration. 

We consider only one protoplanet in our present calculations. 
There is nonetheless a possibility that other protoplanets and even fully formed planets might have existed in the solar nebula at that time.
The presence of other protoplanets would not change our results, since their orbital separation is $\sim10r_{\rm H}\simeq0.2$ au, which is larger than $h_{\rm ch}$. 
The chondrules produced by a protoplanet are accreted only by the protoplanet and surrounding planetesimals. 
Formation of giant planets affects the eccentricities of planetesimals. 
The perturbation from giant planets makes planetesimals dynamically hot.
If the timescale of protoplanet growth becomes longer and the protoplanet does not get its isolation mass within a disk lifetime,
 the mass of accreted chondrules would decrease, as we see in \S \ref{sect:chondrule_acc}. 
In a subsequent paper, we will perform full N-body simulations of planetary growth under the existence of a giant planet,
 and examine the eccentricities of planetesimals and formation of chondrules by impact jetting (S. Oshino et al., in prep).

We have not considered the space and velocity distribution of chondrules and planetesimals in our calculations.
When planetesimals have larger eccentricities and inclinations due to the perturbations from giant planets,
 or chondrules are spatially concentrated by some mechanism such as streaming instability \citep{Youdin&Johansen2007},
 the relative velocity and collisional probability between planetesimals and chondrules are largely changed in an orbit, especially for the vertical direction.
\cite{Guillot+2014} examine how disk turbulence affects the collisional probability between dust particles and planetesimals including their 3D spatial distributions. 
However, accretion of dust particles onto planetesimals taking into account of both their spatial and velocity distributions remains to be explored. 
Meanwhile, our results are not largely changed as long as the picture of oligarchic growth in our fiducial model is not changed.

In \S \ref{sect:mpl}, we perform simulations with $m_{\rm pl,min}<10^{20}$ g for completeness. 
However, the mass of planetesimals strongly affects the onset of runaway growth \citep{Wetherill&Stewart1989,Kokubo&Ida1996,Kobayashi+2016}.
When $m_{\rm pl,min}$ is smaller than a threshold value, planetesimals grow up orderly until certain conditions are satisfied such that runaway growth begins.
Even if runaway growth occurs in a swarm of planetesimals that have $m_{\rm pl,min} < 10^{20}$ g, the mass distribution of planetesimals in oligarchic growth
 would be affected by $m_{\rm pl,min}$ \citep{Morishima2017}.	

It is also important to comment on the isolation mass which regulates the end of chondrule formation in our simulations.
In our fiducial model, the isolation mass of a protoplanet is $1.4M_{\oplus}$. 
Even if $f_{\rm d}=1$, the final mass of protoplanets is larger than the current mass of the asteroid belt.
Such large bodies can be eliminated by the perturbations from giant planets or planetary migration. 
After the giant planets are formed, protoplanets are scattered by their perturbations \citep[e.g.,][]{Petit+2002}.
And, type I migration becomes effective when the protoplanetary mass is larger than $\sim 0.1$-$1M_{\oplus}$
 at 1-3 au \citep[see \S 4.3 in][]{Hasegawa+2016a}.

While we have so far considered the possibility that chondrules formed via impact jetting will be accreted by their surrounding planetesimals,
 it might be interesting to discuss another possibility: formation of planetesimals directly from chondrules ejected from planetesimal collisions. 
This possibility may work well to account for the currently existing meteoritic data \citep{Alexander+2008}.
Unless planetesimal formation from chondrules is not a dominant process, our results would not be largely changed,
 since these planetesimals also produce chondrules by impact jetting.

\section{Conclusions}\label{sect:conclusion}

Investigating the process of chondrule accretion provides us with profound insights into the origins of our Solar System, as well as their formation process. 
When a large number of massive planetesimals are present, which can accrete chondrules, they grow up to be a protoplanet.
The isotope measurement suggested that the timescale of Mars formation is less than the timescale of chondrule formation \citep{Dauphas&Pourmand2011}.
We have investigated chondrule accretion onto a protoplanet and planetesimals in oligarchic growth \citep{Kokubo&Ida1998} using the simple analytical approach.
In our simulations, we consider an impact jetting model as the chondrule formation model. 
When the collision velocity exceeds $2.5 \mbox{ km s}^{-1}$, chondrules are formed via planetesimal collisions \citep{Johnson+2015, Wakita+2016L, Wakita+2016b}. 
The mass of the cumulative formed chondrules is about 1 \% of the protoplanet mass when planetesimal collisions transform about 1 \% the impactor's mass into (the progenitor of) chondrules.
The protoplanet accretes about half of the formed chondrules. 
The other half are accreted by planetesimals. 
In our simulations, we divide planetesimals into 20 mass bins. 
The smallest planetesimal bin has the largest amount of chondrules in all planetesimals, which is about 10\% of the formed chondrules.

We have performed simulations with changing the mass of the smallest planetesimals, the orbital radius, the timescale of gas depletion,
 the efficiency of chondrule accretion by planetesimals, chondrule formation efficiency in the impact jetting model, and chondrule formation models. 
Under the condition that a protoplanet reaches its isolation mass,
 the amount of chondrules accreted by the smallest planetesimals is about 10\% of the formed chondrules for all the runs.
This amount is poorly changed by the chondrule formation models, since it is determined by the timescales of chondrule accretion.
The mass of chondrules accreted by planetesimals is too small to explain the chondrule fraction in chondrites.
Our results indicate that chondrules accreted by planetesimals make a layer on their surfaces. 
Only if chondrites come from this layer, the chondrule fraction in chondrites may be explained. 

\acknowledgments

The authors thank a referee, B. Johnson, for helpful comments.
Numerical computations were carried out on the PC cluster at Center for Computational Astrophysics, National Astronomical Observatory of Japan.
Y. H. is supported by JPL/Caltech.
Part of this research was carried out at the Jet Propulsion Laboratory, California Institute of Technology under a contract with NASA.

\appendix

When a planetesimal stays in $z\leq h_{\rm ch}$, where $z$ is the distance from the midplane,
 the planetesimal can accrete chondrules.
The motion of a planetesimal is given by Hill's equation \citep{Nakazawa&Ida1988}, 
\begin{eqnarray}
	z = i_{\rm pl} r \sin{\left( \Omega_K t - \Omega_{\rm pl} \right)} ,
\end{eqnarray}
where $\Omega_{\rm pl}$ is the longitude of the ascending node of a planetesimal. 
A planetesimal stays in $z\leq h_{\rm ch}$, until 
\begin{eqnarray}
	t \leq \frac{1}{\Omega_K} \mbox{ asin}{\left( \frac{ h_{ch} }{ i_{pl}r }\right)},
\end{eqnarray}
from this passed its ascending node.
The fraction of the timescale that a planetesimal stays in $|z|\leq h_{\rm ch}$ in a orbit ($f_{i_{\rm pl}}$) is given as 
\begin{eqnarray}
	f_{i_{\rm pl}} &=& \frac{ \frac{4}{\Omega_K} \mbox{ asin}{\left( \frac{ h_{ch} }{ i_{pl}r }\right)} }{T_K} = \frac{2}{\pi} \mbox{ asin}{\left( \frac{ h_{\rm ch} }{ i_{\rm pl}r }\right)}.
\end{eqnarray}
This factor is defined when $h_{\rm ch}<i_{\rm pl}r$.

\ \newline

\end{document}